\documentclass[fleqn,usenatbib]{mnras}

\usepackage{newtxtext,newtxmath}

\usepackage[T1]{fontenc}

\DeclareRobustCommand{\VAN}[3]{#2}
\let\VANthebibliography\thebibliography
\def\thebibliography{\DeclareRobustCommand{\VAN}[3]{##3}\VANthebibliography}

\usepackage{graphicx}	
\usepackage{amsmath}	

\usepackage{amssymb}	
\usepackage{makecell}
\usepackage{float}
\usepackage{multirow}
\usepackage{hyperref}
\usepackage{url}
\usepackage[normalem]{ulem}

\usepackage{xspace}
\newcommand{\Jy}{\text{\,Jy}\xspace}
\newcommand{\mJy}{\text{\,mJy}\xspace}

\newcommand{\GHz}{\text{\,GHz}\xspace}
\newcommand{\MHz}{\text{\,MHz}\xspace}

\newcommand{\rad}{\text{\,rad}\xspace}

\newcommand{\radPerSqm}{\text{\,rad\,m}\mbox{$^{-2}$\xspace}}
\newcommand{\mins}{\text{\,min}\xspace}
\newcommand{\hrs}{\text{\,hr}\xspace}
\newcommand{\dys}{\text{\,days}\xspace}
\newcommand{\yrs}{\text{\,yr}\xspace}
\newcommand{\seconds}{\text{\,s}\xspace}
\newcommand{\Msun}{\,M_\odot\xspace}
\newcommand{\kpc}{\text{\,kpc}\xspace}
\newcommand{\pc}{\text{\,pc}\xspace}%
\newcommand{\AU}{\text{\,AU}\xspace}%
\newcommand{\cms}{\text{\,cm}\xspace}
\newcommand\arcdeg{\mbox{$^\circ$}\xspace} 

\newcommand{\muG}{\,\mu\text{G}\xspace}

\newcommand\phn{\phantom{0}}%

\title[V404 Cyg's rapidly evolving polarized jet]{Short Timescale Evolution of the Polarized Radio Jet during V404 Cygni's 2015 Outburst}

\newcommand{\AuthorList}{%
A.~K.~Hughes,$^{1}$\thanks{E-mail: hughes1@ualberta.ca}
G.~R.~Sivakoff,$^{1}$
C.~E.~Macpherson,$^{2}$
J.~C.~A.~Miller-Jones,$^{2}$
A.~J.~Tetarenko,$^{3,4}$\thanks{NASA Einstein Fellow},\newauthor 
D.~Altamirano,$^{5}$
G.~E.~Anderson,$^{2}$
T.~M.~Belloni,$^{6}$
S.~Heinz,$^{7}$
P.~G.~Jonker$^{8,9}$,
E.~G.~K\"ording$^{9}$,
\newauthor
D.~Maitra$^{10}$,
S.~B.~Markoff$^{11,12}$,
S.~Migliari$^{13,14}$,
K.~P.~Mooley$^{15,16}$,
M.~P.~Rupen$^{17}$,
D.~M.~Russell$^{18}$,\newauthor
T.~D.~Russell$^{19}$,
C.~L.~Sarazin$^{20}$,
R.~Soria$^{21,22,23}$, and
V.~Tudose$^{24}$}
\newcommand{\EndAffil}{%
$^{1}$Department of Physics, University of Alberta, CCIS 4-181, Edmonton, AB T6G 2E1, Canada\\
$^{2}$International Centre for Radio Astronomy Research- Curtin University, GPO Box U1987, Perth, WA 6845, Australia\\
$^{3}$Department of Physics $\&$ Astronomy, Texas Tech University, Lubbock, TX 79409-1051, USA\\
$^{4}$East Asian Observatory, 660 N. A`oh\={o}k\={u} Place, University Park, Hilo, HI 96720, USA\\
$^{5}$School of Physics and Astronomy, University of Southampton, Southampton, SO17 1BJ, UK\\
$^{6}$INAF-Osservatorio Astronomico di Brera, via E. Bianchi 46, I-23807 Merate, Italy\\
$^{7}$Department of Astronomy, University of Wisconsin Madison, 475 N. Charter Street, Madison, WI 53706, USA\\
$^{8}$SRON, Netherlands Institute for Space Research, Sorbonnelaan, 2, NL-3584CA Utrecht, the Netherlands\\
$^{9}$Department of Astrophysics/IMAPP, Radboud University, P.O. Box 9010, NL-6500 GL Nijmegen, The Netherlands\\
$^{10}$Department of Physics and Astronomy, Wheaton College, Norton, MA 02766, USA\\
$^{11}$Anton Pannekoek Institute for Astronomy, University of Amsterdam, Science Park 904, NL-1098 XH, Amsterdam, The Netherlands\\
$^{12}$Gravitation Astroparticle Physics Amsterdam (GRAPPA) Institute, University of Amsterdam, Science Park NL-904, 1098 XH Amsterdam, The Netherlands\\
$^{13}$Aurora Technology BV for the European Space Agency, ESAC/ESA, Camino Bajo del Castillo s/n, Urb.~Villafranca del Castillo, 28691, Villanueva de la Ca\~nada, Madrid, Spain\\
$^{14}$Institut de Ci\`encies del Cosmos (ICC), Universitat de Barcelona (IEEC-UB), Mart\'i i Franqu\`es 1, E08028 Barcelona, Spain.\\
$^{15}$National Radio Astronomy Observatory, Socorro, NM 87801, USA\\
$^{16}$Caltech, 1200 E. California Blvd. MC 249-17, Pasadena, CA 91125, USA\\
$^{17}$Herzberg Institute of Astrophysics, National Research Council of Canada, Penticton, BC V2A 6J9, Canada\\
$^{18}$Center for Astro, Particle and Planetary Physics, New York University, Abu Dhabi, PO Box 129188, Abu Dhabi, UAE\\
$^{19}$INAF/IASF Palermo, via Ugo La Malfa 153, I-90146 Palermo, Italy\\
$^{20}$Department of Astronomy, University of Virginia, P.O. Box 400325, Charlottesville, VA 22901, USA\\
$^{21}$College of Astronomy and Space Sciences, University of the Chinese Academy of Sciences, Beĳing 100049, China\\
$^{22}$Sydney Institute for Astronomy, School of Physics A28, The University of Sydney, Sydney, NSW 2006, Australia\\
$^{23}$INAF - Osservatorio Astrofisico di Torino, Strada Osservatorio 20, 10025, Pino Torinese, Italy\\
$^{24}$Institute for Space Sciences, Atomistilor 409, PO Box MG-23, 077125 Bucharest-Magurele, Romania}

\author[A. K. Hughes et al.]{%
\AuthorList
\\%
Affiliations are listed at the end of the paper
}

\date{%
Accepted 2023 January 25. Received 2023 January 25; in original form 2022 October 4}

\pubyear{2023}

\begin{document}
\label{firstpage}
\pagerange{\pageref{firstpage}--\pageref{lastpage}}
\maketitle

\begin{abstract}
We present a high time resolution, multi-frequency linear polarization analysis of Very Large Array (VLA) radio observations during some of the brightest radio flaring (${\sim} 1\Jy$) activity of the 2015 outburst of V404 Cygni. The VLA simultaneously captured the radio evolution in two bands (each with two 1 GHz base-bands), recorded at 5/7\GHz and 21/26\GHz, allowing for a broadband polarimetric analysis. Given the source's high flux densities, we were able to  measure polarization on timescales of ${\sim}13\,$minutes, constituting one of the highest temporal resolution radio polarimetric studies of a black hole X-ray binary (BHXB) outburst to date. Across all base-bands, we detect variable, weakly linearly polarized emission (${<} 1\%$) with a single, bright peak in the time-resolved polarization fraction, consistent with an origin in an evolving, dynamic jet component. We applied two independent polarimetric methods to extract the intrinsic electric vector position angles and rotation measures from the 5 and 7$\,$GHz base-band data and detected a variable intrinsic polarization angle, indicative of a rapidly evolving local environment or a complex magnetic field geometry. Comparisons to the simultaneous, spatially-resolved observations taken with the Very Long Baseline Array at 15.6\GHz, do not show a significant connection between the jet ejections and the polarization state.
\end{abstract}




\section{Introduction}
A black hole X-ray binary (BHXB) is an interacting binary system composed of a stellar-mass black hole accreting material from a companion star. Standard features of BHXBs are jets and winds, making them ideal candidates for the study of accretion-fed outflows. The majority of these systems spend most of their lifetimes in quiescence, accreting small amounts of matter, at low X-ray luminosities ($L_X \lesssim 10^{32}{\rm\,erg\,s^{-1}}$). Most of the known systems sporadically enter into bright ($L_X > 10^{35}{\rm\,erg\,s^{-1}}$), transient outbursts that last weeks to years \citep[e.g.,][]{2016ApJS..222...15T}, allowing for real-time observations of the evolving accretion flow \citep[best measured at X-ray frequencies; e.g., ][]{1999ApJ...519L.159B,2004ApJ...601..439T,2012A&A...538A...5K,2014MNRAS.442.1767P} and relativistic jets \citep[best measured at radio through infrared frequencies; e.g., ][]{2002ApJ...573L..35C,2013MNRAS.436.2625V,2015MNRAS.450.1745R,2015ApJ...805...30T}. 

During an outburst, the morphological evolution of the jet closely correlates with the X-ray properties \citep[i.e., accretion states; e.g.,][]{2006csxs.book..157M,2010LNP...794...53B,2010LNP...794..115F}. In the hard accretion state, an optically thin X-ray corona dominates the X-ray emission and the jet adopts a steady, compact structure. In some systems, the compact jet is observed to persist into quiescence \citep[e.g.,][we note that, for many BHXBs, jets in quiescence will  have flux densities below the detection capabilities of most facilities]{2006MNRAS.370.1351G,2016MNRAS.456.2707P}. Compact jet spectra are described by optically-thick, partially self-absorbed synchrotron emission with an inverted or flat spectral index ($\alpha \gtrsim 0$; with flux density $F_\nu \propto \nu^\alpha$) up to a break frequency (typically in the sub-mm or infrared regime) where the spectra become optically thin ($\alpha \sim -0.6$) to higher frequency emission \citep{2013MNRAS.429..815R}. The flat/inverted spectral index is thought to result from the superposition of multiple spatially-unresolved synchrotron components originating from different positions along the jet axis  \citep{1979ApJ...232...34B}.

Conversely, in the soft accretion state, thermal emission from the accretion disk dominates the X-ray spectrum, and the radio emission from the compact jet decreases or is fully quenched \citep[e.g.,][]{2011ApJ...739L..19R,2020MNRAS.498.5772R}. During the hard-to-soft transition, one or more blobs of discrete jet ejecta are typically launched, and these ejecta have been spatially resolved in several sources \citep[e.g.,][]{1994Natur.371...46M,1995Natur.375..464H,2001ApSSS.276...45H,2017MNRAS.468.2788R,2019Natur.569..374M}. 

The ejection events are attributed to brief periods of highly efficient plasma production at the base of the jet, creating (often adiabatically) expanding plasma knots threaded with complex magnetic fields \citep[e.g., the van der Laan ---vdL--- model,][and references therein]{1966Natur.211.1131V,1988ApJ...328..600H,1995xrbi.nasa..308H}. Each ejection has an emission spectrum characterized by a single self-absorbed synchrotron source with a temporally evolving electron/lepton population. As the ejection propagates and expands, the self-absorption turnover transitions to lower frequencies, extending deep into the radio regime and resulting in an observing bandwidth that is optically thin in its entirety \citep[e.g.,][]{2014MNRAS.437.3265C,2015MNRAS.451.3975C,2020MNRAS.491L..29W}. In the radio, these ejections are observed as multi-frequency flares with well-defined rise and decay phases that last minutes to days. Due to the expansion-driven, evolving optical depth, the lower frequency components are broadened in time and temporally delayed with respect to the higher frequency counterparts \citep{1998A&A...330L...9M}. Flaring events from Active Galactic Nuclei (AGN, the large-scale analogous of BHXBs) have also been modelled based on the adiabatic expansion of jet plasma \citep[]{2008ApJ...682..361Y,2009A&A...496...77F,2009A&A...508L..13M,2021ApJ...917....8B,2021ApJ...923...54M}.

Alternative flaring models can also be applied to BHXB observations, such as the ``shock-in-jet'' picture that is typically associated with AGN \citep[e.g.,][]{1985ApJ...298..114M,2001MNRAS.325.1559S}. Within this framework, each flare is the result of shocks within a (quasi-) steady jet accelerating particles and temporarily enhancing emission intensities \citep[e.g.,][]{2004MNRAS.355.1105F,2004A&A...415L..35T,2011MmSAI..82..104T,2013MNRAS.429L..20M}.  

Most radio jets from BHXBs are described by their photometric, spectral, and (when available) spatial properties. However, a much smaller fraction of studies explore the linear polarization that results from a synchrotron dominated emission spectrum. 
For optically thin and optically thick synchrotron emission, the maximum expected linear polarization fraction is $f_\lambda = (3p+3)/(3p+7)\times100\,{\%}\approx70\,{\%}$ and $f_\lambda = 3/(6p+13)\times100\,{\%}\approx10\,{\%}$,  respectively \citep[assuming a uniform magnetic field and adopting a typical value for the electron energy distribution index, $p=2.2$;][]{1969ARA&A...7..375G,2011hea..book.....L}. Complex or evolving magnetic fields, disadvantageous lines of sight, Faraday depolarization, and the superposition of multiple components are a few mechanisms that can depolarize the observed radio emission. BHXBs with polarimetric radio analyses typically have linear polarization fractions $\lesssim10\%$ with a rare few reaching $\sim50\%$ \citep[e.g.,][]{1992ApJ...400..304H,2000ApJ...540..521H,2003Ap&SS.288...79F,2007MNRAS.378.1111B,2013MNRAS.432..931B,2014MNRAS.437.3265C,2015MNRAS.451.3975C}. The polarization fraction measures how ``ordered'' the local magnetic field is (or appears to be), while the direction of the observed electric vector position angle (EVPA) is a measure of the local absorption conditions, jet position angle, magnetic field orientation, as well as Faraday rotation between the emission and the observer. After measuring and removing the effect of the Faraday rotation, the derived intrinsic EVPA can provide an indirect measure of the jet orientation \citep[e.g.,][]{2014MNRAS.437.3265C,2015MNRAS.450.1745R}. In cases where polarized emission is combined with spatially resolved, total intensity observations, polarimetry can directly probe the underlying magnetic field strength and orientation \citep[e.g.,][]{2004MNRAS.354.1239S}.

Despite the established observational relationship between the X-ray and radio properties (i.e., the accretion flow and relativistic jet), the physical mechanisms responsible for the launching and evolution of jets are yet to be fully understood. Most theories recognize that the local magnetic fields (and their disk/black-hole interactions) play an essential role in extracting energy from the black hole/accretion disk \citep[e.g.,][]{1977MNRAS.179..433B,1982MNRAS.199..883B} and the initial launching and collimation of relativistic jets \citep[e.g.,][]{2004ApJ...605..656V,2007MNRAS.380...51K,2010MNRAS.402....7M}. These highly energetic processes can leave imprints on the evolving magnetic fields, making time-resolved radio polarimetry, particularly around BHXB ejection events, a valuable (yet underutilized) tool. Outbursts from BHXBs occur at a moderate frequency \citep[e.g., several times a year;][]{2016ApJS..222...15T}, with a rare subset (e.g., once per decade) achieving X-ray luminosities near (or exceeding) the Eddington luminosity, and Jansky level radio flux densities \citep[e.g., V404 Cygni's 1989 outburst reached 1.6\,Jy at 4.9\, GHz;][]{1996A&A...309..781O,1992ApJ...400..304H}. During highly luminous outbursts, we can study accretion and accretion-rooted phenomena with extraordinary levels of detail, capturing, in real-time, jet ejections at flux densities that allow for a refined spectral and temporal resolution for both total intensity and polarimetric observations. On 2015 June 15, the BHXB V404 Cygni (henceforth V404 Cyg) began one of these rare outbursts. 

\subsection{V404 Cygni}
First discovered in 1989, V404 Cyg (also known as GS 2023+338) is a low-mass transient BHXB that has undergone four recorded outbursts. Of the four outbursts, two were caught in real-time; the initial discovery with the Ginga satellite \citep{1989IAUC.4782....1M} and the most recent outburst discovered by the Burst Alert Telescope aboard the {\em Neil Gehrels Swift Observatory} \citep{2015GCN.17929....1B}. Searches through historical photo plates identified that there were additional outbursts in 1938 and 1956 \citep{1989IBVS.3362....1R}. Observations of the main-sequence companion star revealed an orbital period of $6.4714 \pm 0.0001$ days and a binary mass function of $f = 6.08 \pm 0.06\Msun = M_{\rm BH}^3 \sin^3 i / (M_{\rm BH}+M_{\rm donor})^2$, where $M_{\rm BH}$ and $M_{\rm donor}$ are the masses of the black hole and donor, respectively, and 
$i$ is the orbital inclination angle \citep{1994MNRAS.271L...5C}. The K spectral type of the companion star, coupled with near-infrared spectroscopy (and modeling of the H-band ellipsoidal modulations), infer a BH mass of $9.0_{-0.6}^{+0.2}\Msun$ with a best fit orbital inclination angle of $67^{+3} _{-1}\arcdeg$ \citep{2010ApJ...716.1105K}. The modelled orbital inclination angle assumes that the optical light curve of the companion star has ${\lesssim} 7\%$ contamination from accretion disk (or jet) emission. However, V404 Cyg has exhibited optical variability in quiescence \citep[e.g., ][]{2003ApJ...582..369Z, 2016ApJ...818L...5B} and, as a result, may have a larger contamination fraction, larger inclination angle, and smaller black hole mass. On the other hand, narrow emission lines suggest that V404 Cyg has a low inclination angle $i<40\arcdeg$ \citep{1993MNRAS.265..834C} and a higher black hole mass. This uncertainty suggests the mass of the black hole is not yet accurate at the 2--7\% precision level quoted above. High angular resolution radio parallax measurements determined a source distance of $2.39\pm0.14\kpc$ \citep{2009ApJ...706L.230M}, making it one of the closest known BHXBs and a superb laboratory for the study of accretion physics. 

During the 1989 outburst, \citet{1992ApJ...400..304H} monitored the radio emission of V404 Cyg between 1989 May 30 and 1991 May 31. The monitoring began when the radio light curves were dominated by the tail of a rapidly decaying (decay timescales of ${\sim} 5\dys$) ``major synchrotron bubble event". At later times, the radio light curves were dominated by a slowly-decaying, nonthermal, optically thick source (e.g., a compact jet) that lasted hundreds of days. Linear polarization was detected for the first 50 days of observations, except for the first observation on 1989 May 30 which did not include adequate polarization calibration. During the decay of the synchrotron bubble, the polarization fraction was a few tenths of a percent, before increasing to a few percent during the period when the slowly-decaying component dominated the radio emission. 

On 2015 June 15, V404 Cyg entered its fourth recorded outburst, and a follow-up campaign showed bright multi-wavelength flaring activity in radio through X-ray wavelengths \citep[e.g.,][]{2015ATel.7658....1M,2015ATel.7665....1M,2015ATel.7666....1M,2015ATel.7661....1T,2015ATel.7708....1T,2016MNRAS.459..554G,2017ApJ...851..148M}, and rapid ($\sim15\,$s) transitions between accretion states \citep[see,][and references therein]{2020A&A...634A..94K}. With radio-through-optical flux densities reaching  ${\sim}\Jy$ levels, V404 Cyg became the brightest BHXB outburst observed in the last decade, characteristic of a high, near-Eddington accretion rate, and its close proximity. The source remained in outburst until the end of June, from which it began decaying, eventually reaching quiescence in mid-August \citep[with the source having a  brief period of renewed activity in 2015 December through 2016 January;][]{2017ApJ...834..104P,2017MNRAS.465L.124M}.

The MASTER Global Robotic Net detected three linear polarization ``events'' using their optical telescope network \citep{2016ApJ...833..198L,2019NewA...72...42L}. In both events, the source exhibited a significant increase in the linear polarization fraction, following a (total intensity) flare, followed by a rapid decrease in the linear polarization fraction during the rise of another flare. The authors favoured a model where decreased X-ray irradiation of the secondary also decreased its optical brightness. In turn, this  makes it easier to detect the polarized non-thermal emission from the jet. The authors favoured this model after having discarded the potential that the jet orientation varied on timescales of tens of minutes; however, a rapid, variable jet orientation was later confirmed \citep[see below and ][]{2019Natur.569..374M}. 

\citet{2016MNRAS.463.1822S}, detected another linear polarization flare using observations with the Nordic Optical Telescope. This flare occurred during a steady rise of optical flux, and preceded some of the brightest optical flaring of the entire outburst. Moreover, the flare preceded the start of a bright radio flare. These authors proposed that the increase in linear polarization could result from multiple ejecta collisions establishing a dominant magnetic field direction perpendicular to the jet axis, and may be the signature for the birth of the ejection that produced the subsequent radio flare. 

During this same outburst, modeling of radio-through-sub-mm observations \citep{2017MNRAS.469.3141T} and Very Long Baseline Array (VLBA) observations \citep{2019Natur.569..374M} uncovered short time-scale flaring of the jet. Here, the jet ejecta account for most, if not all, of the observed flaring. However, we note that detailed modelling of the X-ray emission suggested that the source may have been continuously accreting at an Eddington accretion rate during the brightest phase of the 2015 outburst \citep{2017MNRAS.465L.124M}. As a result, the jet ejecta in V404 Cyg are not clearly associated with the same discrete-jet launching process that occurs during hard-to-soft state transitions in BHXBs, which occur around lower accretion rates \citep{2004MNRAS.355.1105F}. The VLBA observations directly resolved several of these ejection events on top of a continuous (but variable) emission from an empirically defined unresolved radio core. The position angle (PA) of the ejecta varied rapidly with time, a phenomenon that was attributed to the Lense-Thirring precession of the accretion disk \citep{2019Natur.569..374M}. Although the authors constrained the period to less than 2.6$\,$hr hours, the rapidly varying PAs suggested that the true period was substantially shorter.

In this paper, we add to the detailed radio analysis of the 2015 outburst detailed in \citet{2017MNRAS.469.3141T} and \citet{2019Natur.569..374M}. Here, our primary focus is the extraction and analysis of V404 Cyg's (radio) polarization properties --- derived from National Science Foundation's Karl G.\ Jansky Very Large Array (VLA) observations --- during some of the outburst's brightest flaring activity on 2015 June 22. The remainder of this paper is structured as follows; in Section \ref{sec:obs}, we introduce our observation and analysis procedure, while in Sections \ref{sec:results} and \ref{sec:discussion}, we present and discuss our results. Finally, we summarize our findings in Section \ref{sec:conclusions}.

\section{Observations and Analysis}
\label{sec:obs}
\subsection{VLA Data Reduction}
The details of the primary VLA observations were first discussed in \citet{2017MNRAS.469.3141T}. V404 Cyg was observed with the VLA (Project Code: 15A-504) on 2015 June 22 with scans on source between 10:37:24 and 14:38:39 UTC in the 4--8\GHz and 18--26\GHz bands. All observations were made with an 8-bit sampler, comprised of two base-bands, with eight spectral windows of sixty-four 2\MHz channels each, giving a total (unflagged) bandwidth of 1.024\GHz per base-band. Henceforth, we will refer to each base-band by its characteristic frequency values of (${\sim}$) 5, 7, 21, and 26\GHz. 

The array was in its most extended A configuration, and was split into two sub-arrays of 14 (sub-array 1) and 13 (sub-array 2) antennas. Sub-array 1 observed the sequence (5/7\GHz)-(21/26\GHz)-(5/7\GHz), while sub-array 2 observed the sequence (21/26\GHz)-(5/7\GHz)-(21/26\GHz). Both sub-arrays cycled between V404 Cyg, observed for 88\seconds per cycle flanked by 32\seconds observations of a nearby gain calibrator (J2025$+$3343). A second epoch was observed during the source's return to quiescence, taken on 2015 July 2, with scans on source from 10:31:08 to 14:01:32 UTC. The observing bands and sub-array schemes remained consistent with the primary June 22 observations \citep{2019MNRAS.482.2950T}. We also analyzed 5 epochs taken between July 11 and August 5, during the source's return to quiescence \citep[Project Code: SG0196; ][]{2016MNRAS.456.2707P}, although we were unable to detect any polarized signal in these latter observations (see Section \ref{sec:rm_pol}). 

We applied standard flagging and calibration to the Stokes $I$ (i.e., total continuum flux density) data using the Common Astronomy Software Application package (\textsc{casa} v5.6; \citealt{2007ASPC..376..127M}). We used 3C48 (0137$+$331) as a flux and absolute (linear) polarization angle calibrator, J2025$+$3343 as a complex gain (aka phase) calibrator, and J2355$+$4950 as an unpolarized leakage calibrator for both sub-arrays. Due to V404 Cyg being weakly polarized, we grouped our polarization calibration solutions on 16\MHz (8 channel) intervals. For our Stokes $I$ flux calibration model, we used the default \textsc{casa} model repository \citep{2017ApJS..230....7P}. However, the standard calibration routine for Stokes $Q$ and $U$ (i.e. linearly polarized flux densities) assumes that the polarization calibrators are point sources. Since we observed with the VLA in its most extended configuration, 3C48 was resolved. The ``degree'' of resolution ranges from a slightly extended Gaussian at 5/7\GHz to multiple distinct components at 21/26\GHz. As a result, we constructed a spatially resolved model image for each observing band. Our model contained information on all four Stokes parameters, assuming no circular polarization, and adopted the spatial distribution of the Stokes $I$ repository models. A detailed description of our polarized model image can be found in Appendix \ref{sec:calibrator_model}. We note that the spatial distribution of the flux densities may differ between Stokes $I$, and Stokes $Q$/$U$ (i.e. linearly polarized flux densities), and, as a result, our measured polarizations are susceptible to systematic calibration errors; in particular, for the 21/26\GHz basebands, where our calibrator is significantly resolved.

\subsection{Imaging}

Since V404 Cygni was expected (and found) to be unresolved regardless of the chosen visibility weighting, we applied a \textit{natural} $uv$-weighting scheme to all of our images, maximizing sensitivity. Moreover, since we also knew that the Stokes $I$ flux density was rapidly changing, we generated our analysis images in Stokes $I$, $Q$, and $U$, on short timescales of 12 or 14 \mins (6 or 7 scans; i.e., ${\sim} 8$ or $9.5 \mins$ on source)\footnote{We made the time bins a variable integer number of scans to avoid combining scans from different sub-arrays. We note that the ${\sim} 20\%$ difference in on-source time has a negligible effect on the analysis.}. These timescales, some of the shortest ever used in a radio polarimetric analysis of a BHXB, balance cadence and polarized sensitivity.

We used the \textsc{wsclean} package \citep{offringa-wsclean-2014} to make all of our polarimetric images. We imaged each base-band independently, as well as Stokes $I$ separately from $Q$ and $U$. In each base-band for every time-bin we had  \textsc{wsclean} output a set of images across a user-set number of channels, as well as a single ``multi-frequency-synthesis'' (MFS) image that stacks all the individual channels. We measured the linear polarization intensities ($P=\sqrt{Q^2 +U^2}$) for each base-band/time-bin pair from the MFS images.

Any observed EVPA at an arbitrary observing wavelength $\lambda$, is related to the intrinsic EVPA, $\chi_0$, through the linear relationship, $\chi(\lambda) = \chi_0 + \text{RM} \cdot\lambda^2$. The slope (i.e., the rotation measure, RM) quantifies the wavelength-dependent \textit{Faraday rotation} of an EVPA due to linearly polarized light propagating through a magneto-ionic plasma. Since our observables are $\chi$ and $\lambda$, the largest detectable rotation measure is inversely proportional to the $\lambda^2$ channel spacing. The linearly-spaced frequency channels result in a $\lambda^2$ channel density that increases with increasing central frequency. To avoid potential biasing of results by the higher frequency observations, we scaled the imaging frequency bins used for rotation measure analysis to maintain a (roughly) constant $\lambda^2$ channel spacing; this resulted in a frequency-space channelization of 16\MHz for the 5\GHz baseband, and 64\MHz for the 7\GHz baseband. Due to their large temporal delays (${\gtrsim30} \mins$) with respect to the 5/7\GHz base-bands, we chose to omit the 21/26\GHz base-bands from the rotation measure analysis. The omission will minimize the overlap of optically thick and optically thin emission, as well as any overlap of emission from different jet components (see Appendix \ref{sec:appendix_delays} for a more comprehensive motivation behind the omission). As a result, we did not scale the frequency binning any broader than 64\MHz.

The larger Stokes $I$ flux densities allowed us to image the total flux density light curves on much shorter timescales (${\sim} 10\seconds$) than is  required for accurate polarimetry. For each spectral window, we produced a high time-resolution light curve, using the publicly available\footnote{\url{https://github.com/Astroua/AstroCompute_Scripts}} imaging scripts detailed in \citet{2017MNRAS.469.3141T}. These images critically allow us to compare the simultaneous Stokes $I$ flux density, and linear polarization evolution. 

We observed an elevated rms noise in each image when compared to the predicted values (see Table \ref{tab:image_params} for a summary). These effects are most significant in the Stokes $I$ images and appear to worsen at higher central frequencies and when larger frequency ranges are used to create a single image. Therefore, we implemented a phase self-calibration routine to explore if the elevated Stokes $I$ noise is biasing the polarimetric results. Our self-calibration routine was broken into three steps that refined the phase calibration solutions on progressively shorter timescales: first, half the length of a source scan, 44\seconds; then a quarter, 22\seconds; and ending with solutions on the integration timescale, 2\seconds. We excluded amplitude self-calibration due to the known Stokes $I$ variability within our imaging intervals. Although the phase self-calibration improved the Stokes $I$ rms noise, we were unable to reach the theoretical limit expected from thermal noise. This result is not unexpected: (i) we are averaging over variable emission (spectrally and temporally) during our imaging routines; (ii) the reduction in baseline coverage due to the division into sub-arrays coupled with the bright emission is expected to limit the dynamic range; (iii) completely automated self-calibration, like we employ, can have difficulties achieving high dynamic ranges; (iv) we only image a ${\approx}51\arcsec\times51\arcsec$ field-of-view, and there can be some added noise across the entire image due to our nearby phase calibrator (approximately 16.6\arcmin from V404 Cyg) --- our primary beams range from 1.6--8.9\arcmin, leading to noise that would be stronger in our lower frequency basebands. Since the self-calibration and its reduction of the Stokes $I$ rms had a negligible effect on both the noise of the Stokes $Q$ and $U$ images and our measured polarimetric parameters, we are confident that the elevated noise is not a significant issue. For the remainder of this analysis, we have adopted our self-calibrated results. 

\begin{table*}
  \centering
  \caption{Table of imaging properties. The highlighted frequency parameters for each base-band are the central frequency of the lowest ($\nu_i$) and highest ($\nu_f$) channels, in addition to the imaging bandwidth ($\Delta\nu$) assuming a typical ${\sim} 15\%$ loss during flagging and calibration. $\Delta t$, is the average time on source. The theoretical rms noise ($\sigma_\text{rms}$) and the median rms noise for each Stokes parameter ($\sigma_I$, $\sigma_Q$, $\sigma_U$) are also highlighted. The high time-resolution images ($\Delta\nu\sim110\MHz$) were excluded from the self-calibration procedure due to the number of images (${\sim} 45000$). The theoretical noise estimates were calculated using the VLA exposure calculator; \url{obs.vla.nrao.edu/ect/}.}\label{tab:image_params}  
  \begin{tabular}{ccccccccc}
  \Xhline{3\arrayrulewidth}
    Base-band  & $\nu_i\,$(MHz) & $\nu_f\,$(MHz) & $\Delta\nu$ (MHz) & $\Delta t$ (s)& $\sigma_\text{rms}$ (mJy) & $\sigma_I$ (mJy) & $\sigma_Q$ (mJy) & $\sigma_U$ (mJy) \\
  \Xhline{3\arrayrulewidth}
  \multirow[t]{3}{*}{\phn 5\GHz} & \multirow[t]{3}{*}{\phn4738} & \multirow[t]{3}{*}{\phn5762} & 850 & 520 & 0.03 & \phn0.2\phn & 0.05 & 0.05  \\ 
   & & & \phn16 & 520 & 0.2\phn & \phn0.4\phn & 0.3\phn & 0.3\phn \\
   & & & 110 & \phn10 & 0.6\phn & \phn2\phantom{.00} & --- & ---   \\
   \multirow[t]{3}{*}{\phn7\GHz} & \multirow[t]{3}{*}{\phn6938} & \multirow[t]{3}{*}{\phn7962} & 850 & 520 & 0.03 & \phn0.3\phn & 0.06 & 0.06 \\
   & & & \phn64 & 520 & 0.1\phn & \phn0.4\phn  & 0.14 & 0.14 \\ 
   & & & 110 & \phn10 & 0.6\phn & \phn3\phantom{.00} & --- & ---  \\
  \multirow[t]{2}{*}{21\GHz} & \multirow[t]{2}{*}{20288} & \multirow[t]{2}{*}{21312} & 850 & 520 & 0.09 & \phn1\phantom{.00}  & 0.14 & 0.15 \\ 
   & & & 110 & \phn10 & 1.5\phn & 10\phantom{.00} & --- & --- \\
    \multirow[t]{2}{*}{26\GHz} &  \multirow[t]{2}{*}{25388} &  \multirow[t]{2}{*}{26412} & 850 & 520 & 0.08 & \phn1.7\phn & 0.3\phn  & 0.2\phn \\
   & & & 110 & \phn10 & 1.5\phn & 12\phantom{.00} & --- & --- \\
  \Xhline{3\arrayrulewidth}
  \end{tabular}
\end{table*}

\subsection{Flux Density Extraction}
We measured the Stokes $I$ flux densities and linear polarization intensities (from the MFS images) from an image plane analysis using the \textsc{casa} task \texttt{imfit}. We fit an elliptical Gaussian component in a small sub-region around the source, fitting for the position, flux density, and shape of the component. Due to the source's weakly polarized emission, at fine spectral resolutions (e.g., the 16$\,$MHz channelization), the Stokes $Q$ and $U$ flux densities are similar in magnitude to (or weaker than) the local peaks in the rms noise (see Appendix \ref{sec:appendix_image}). Often our attempts to freely fit the Stokes $Q$ and $U$ images using \texttt{imfit} did not converge or converged on artificial noise signals. Therefore, we decided to fix the shape of the component, and only fit for the flux density in the region (i.e., we performed forced aperture photometry). We set the component shape to be the synthesized beam of each image, and used the position of the $P$ peak (for each time bin and base-band) as the position of our aperture. We extracted the rms of each image using a large annular region centred on the source. To check for bias by a non-zero background we subtracted the mean flux density in the rms region from the flux density of the source. The background subtraction had a negligible effect on our results.

The fine (spectral) resolution images uncovered anomalous channels (${\sim} 1 \mbox{--} 2$ per time bin) that were missed during flagging and calibration, or corrupted during imaging. We apply a $\sigma$-clipping routine to remove these channels from the Stokes $I$ spectrum of each time bin. After constructing a model spectrum by passing our Stokes $I$ data through a narrow Gaussian filter with $\sigma=2.5$ data points, corresponding to 5 and 20\MHz at 5 and 7\GHz, respectively, any flux density point that was ${>}{\,}3$ residual standard deviations from the model spectrum was flagged. We continued the routine until the fractional difference in residual standard deviations between the current and previous iteration was ${\leq}\,0.001\%$. The channels removed from the total intensity spectra were recorded and subsequently removed from the $Q$ and $U$ spectra. No further data manipulation was applied.

\subsection{Polarization Properties}
We derived all polarization properties from the flux densities extracted during image plane analysis. The polarization intensity images, $P_\lambda = \sqrt{Q_\lambda^2 + U_\lambda^2}$,  for an image with a central wavelength $\lambda$, were created from the Stokes $Q$ and $U$ images using the native \textsc{casa} task \texttt{immath}. Since $P_{\lambda}$ is positive definite, we debiased each polarization intensity using the correction from  \cite{2012PASA...29..214G}; $P_{\lambda,0} = \sqrt{P_\lambda^2 - 2.3\sigma_{QU}^2}$. To remain consistent with the RM synthesis routine (Section \ref{sec:rmsynth}), we have chosen $\sigma_{QU} \equiv \frac{1}{2}(\sigma_Q + \sigma_U)$ to parameterize the noise in $P_{\lambda}$, noting that $\sigma_Q\,{\approx}\,\sigma_U$ for all of our images. The polarization fraction adopts its standard definition, $f_\lambda \equiv P_{\lambda,0}/I_\lambda$, and we approximated its error using Gaussian error propagation. We recognize that the MFS images will experience a degree of bandwidth depolarization due to averaging over an intra-band Faraday rotation. However, at our detected rotation measures ($\text{|RM|} \sim 100\radPerSqm$), even at the lowest frequencies, the amount of depolarization is insignificant; $\Delta f_\lambda/f_\lambda \lesssim 1\%$. 

To extract the intrinsic EVPA and rotation measure from each time bin, we applied two independent methods: rotation measure synthesis and a custom Markov-Chain Monte Carlo (MCMC) routine. Meaningful RM synthesis results requires a band-averaged, polarized S/N of $P_{\lambda,0}/\sigma_{QU} \gtrsim 7$ \citep[e.g.,][]{2005A&A...441.1217B,2012ApJ...750..139M}. To ensure the significance of each detection, we enforce the $P_{\lambda,0}/\sigma_{QU} > 7$ restriction on the 5/7\GHz base-bands separately. Our aggressive restriction was motivated by the susceptibility of weakly polarized data to spurious effects from imperfect leakage calibration. As a result, we limited the intrinsic EVPA and rotation measure analysis to the 13 time bins between 11:15 and 13:53 UTC. Data tables including our polarimetric measurements can be found in Appendix \ref{sec:data_tables}.

\subsubsection{Rotation Measure Synthesis}
\label{sec:rmsynth}
Rotation measure synthesis derives the linear polarization parameters of a source through its structure(s) in \textit{Faraday space}; i.e., its Faraday dispersion function \citep[FDF; see,][for a comprehensive description]{1966MNRAS.133...67B,2005A&A...441.1217B,2012ApJ...750..139M,2012MNRAS.424.2160H}. We generated each FDF using the \textsc{rm-tools}\footnote{\url{https://github.com/CIRADA-Tools/RM-Tools}} repository, currently developed and maintained by the Canadian Initiative for Radio Astronomy Data Analysis (CIRADA). To mitigate any aliasing at large rotation measures, we fixed the FDF domains at $\pm1.5\times 10^5\radPerSqm$, i.e., twice the rotation measure that corresponds to a ${\sim} 50\%$ drop in sensitivity at our spectral channelization. Furthermore, we fixed the bin size at $75\radPerSqm$, a factor of 20 (twice the median polarized S/N) smaller than the full width at half maximum of the rotation measure synthesis function. The package typically quantifies the noise in each FDF ($\sigma_\text{RM}$) using the median absolute deviation after masking the strongest rotation measure component. We chose to use the rms noise as it was a factor of ${\sim} 2$ larger, and thus, increased our confidence in each detection. Any FDF component that satisfied a ${>}\,5\sigma_\text{RM}$ condition was recorded. During the construction of each FDF, the observed EVPAs are de-rotated to their values at the weighted mean of the $\lambda^2$ channels, with a $1/\sigma_{QU}^2$ weighting. The intrinsic EVPA is calculated from a further de-rotation using the best-fit rotation measure; i.e., $\chi_0 = \chi_w - \text{RM}\cdot\lambda_w^2$, where $\lambda^2_w$ is the weighted average of all $\lambda^2$ channels and $\chi_w$ is the observed polarization angle at $\lambda^2_w$.

\subsubsection{MCMC}
Since V404 Cyg is weakly polarized, we also employ a simple Bayesian forward model to fit the polarization parameters directly to the Stokes fluxes. Consistency between the two methods is an important check to mitigate the potential that our derived polarization parameters originate from noise, as opposed to an intrinsic signal. Our fitting functions adopt the following forms; 

\begin{align}
    &\widetilde{Q}_\lambda = \widetilde{I}_\lambda \widetilde{f}_{\lambda}\cos\left(2\chi_w + 2\text{RM} \cdot(\lambda^2-\lambda_w^2)\right)\text{; and} \\
    &\widetilde{U}_\lambda = \widetilde{I}_\lambda \widetilde{f}_{\lambda}\sin\left(2\chi_w + 2\text{RM} \cdot(\lambda^2-\lambda_w^2)\right).
\end{align}
  We chose to fit for $\chi_w$, to remain consistent with the RM synthesis routine. The superscript, $\lambda$, in equations (1) and (2) denotes the central wavelength of the spectral channel of interest. The model parameters for Stokes $I$ ($\widetilde{I}_\lambda$) and the linear polarization fraction ($\widetilde{f}_\lambda$) were excluded from the fitting procedure, due to negligible correlation with the quantities of interest (RM and $\chi_w$). Instead, the Stokes $I$ and polarization fraction models were smoothed using a Savitzky-Golay filter \citep{citeulike:4226570}, retaining the overall structure while removing stochastic variability, and stabilizing the fitting routine. 

We assumed the sampled flux densities were independently distributed normal random variables, resulting in a log-likelihood function ($\mathcal{L}$) of the following form, 
\begin{multline}
    \log\mathcal{L} = -\sum_{\lambda}\left[\log\sqrt{2\pi\widetilde{\sigma}^2_{Q,\lambda}} + \frac{(Q_\lambda-\widetilde{Q}_\lambda)^2}{2\widetilde{\sigma}_{Q,\lambda}^2} \right. \\ + \left.\log\sqrt{2\pi\widetilde{\sigma}^2_{U,\lambda}} + \frac{(U_\lambda-\widetilde{U}_\lambda)^2}{2\widetilde{\sigma}^2_{U,\lambda}}\right],
\end{multline}
where $Q_\lambda$/$U_\lambda$ and $\widetilde{Q}_\lambda$/$\widetilde{U}_\lambda$ are the measured and modelled flux densities, respectively. We added two additional modeling parameters, $\sigma_{Q,\text{sys}}$ and $\sigma_{U,\text{sys}}$, that are channel independent variances to account for missed systematic effects. The variances seen in equation (5) are the sum of the measured rms noise variance and our systematic addition (e.g., $\widetilde{\sigma}^2_{Q,\lambda}\equiv \sigma^2_{Q,\lambda} + \sigma^2_{Q,\text{sys}}$).

We used the Markov-Chain Monte Carlo algorithm implemented through Python's \textsc{emcee} package. \textsc{emcee} is a pure-Python implementation of Goodman and Weare's Affine Invariant Markov chain Monte Carlo Ensemble Sampler \citep{2013PASP..125..306F,2010CAMCS...5...65G}; a modified version of the classic Metropolis-Hastings algorithm, simultaneously evolving a select number of \textit{walkers} through parameter space. The number of (sampling) walkers was fixed at five times the number of dimensions, 20. We chose four broad, uniform, and uninformative priors to reflect the lack of \textit{a priori} information on V404 Cyg's polarization state. The systematic variance priors were positive definite, with maximum values chosen to be twice the variance of the measured flux densities. The rotation measure prior adopted the FDF domain, $\pm1.5\times10^5\radPerSqm$. A uniform prior was unable to capture the circularity of the EVPA. As a result, individual walkers frequently would become trapped in the local minima created by the prior's edges, subsequently inhibiting convergence. To combat this, we expanded the prior to $\pm 3\pi/2\rad$, while maintaining the initial condition distribution for the physically meaningful range of $\pm\pi/2\rad$. We initialized each run with 80 walkers, four times the number intended for sampling. Following an initial set of ``burn-in" iterations, we removed the 60 walkers with the lowest posterior probabilities and adopted the remaining 20 as the starting positions for sampling. After sampling, we verified that each simulation converged by visually inspecting the walkers over a large number of autocorrelation times.

We adopted the median of each posterior distribution as the best-fit value of our model, and the ranges between the median and the $15^\text{th}$/ $85^\text{th}$ percentiles as the 1$\sigma$ $(-)/(+)$ uncertainties, noting that the measured uncertainties are purely statistical. Once again, the intrinsic EVPA was solved for using, $\chi_0 = \chi_w - \text{RM}\cdot\lambda_w^2$, and we calculated its error using standard Gaussian error propagation.

\section{Results}
\label{sec:results}
By splitting our ${\sim} 3.5\hrs$  observation into sixteen ${\sim} 13 \mins$ time bins, we have measured the temporal evolution of the linear polarization fraction (Figure~\ref{fig:frac_pol}), rotation measure, and intrinsic EVPA  (both Figure~\ref{fig:rm_pol}) during the 2015 June 22 flaring events of V404 Cyg. In this section, we present our polarimetric results. We note that weak linear polarization fractions should be treated with caution; in Appendix \ref{sec:phase_cal_evolution}, we compare our results to the simultaneous evolution of the phase calibrator ($f_\lambda \sim 2\%$) to ensure that significant changes we see in V404 Cyg arise from physical evolution, and not systematic calibration effects.

\subsection{Linear Polarization Fraction}
\label{sec:f_pol}

Each base-band showed a weak but variable degree of linear polarization with a maximum linear polarization fraction that decreased with decreasing frequency; i.e, maxima of ${\sim}$0.22, 0.25, 0.5, 0.75${\%}$ for the 5, 7, 21, and 26\GHz base-bands, respectively. The maximum linear polarization fraction occurs between the peaks of the first and second Stokes $I$ flare, and, like Stokes $I$, occurs at later times for lower-frequency observations. There is evidence of a second (much weaker) linear polarization fraction peak in the 21$\,$GHz base-band, between the second and third flare (at ${\sim}$ 12:50 UTC). This secondary peak is marginally detected in the 5/7\GHz base-bands, but is consistent with noise at 26\GHz. Additionally, in the 21/26\GHz base-bands, at late times the linear polarization fraction begins to increase alongside the decay of the third flare. A similar increase is not observed in the 5\GHz base-band, with a marginal trend seen in 7\GHz, although temporal delays would have likely shifted any peak at these frequencies beyond our observing time. 

In the 2015 July 2 observations, during V404 Cyg's return to quiescence, the Stokes $I$ flux densities had decreased to $\sim4\,$mJy across all base-bands. As a result, we are unable to detect weakly polarized emission, and the source showed no polarization with a 99$\%$ confidence upper limit on the polarization fraction of 1.0, 0.9, 2.2, 2.4\% for the 5, 7, 21, and 26$\,$GHz base-bands, respectively. Here we calculated upper limits following  \citet{2006PASP..118.1340V}. Furthermore, we analyzed 5 subsequent epochs of the 5/7$\,$GHz observations taken between 2015 July 15 and 2015 August 5 \citep[see,][for details]{2017ApJ...834..104P}. Of these 5 epochs, data on 2015 August 5 had the most constraining upper limits, 6.0, and 5.1$\%$ (for the 5 and 7$\,$GHz base-bands respectively), with all other epochs having upper limits between $10-25\%$. While we cannot detect a weakly polarized signal with these observations, a $\sim 5\%$ limit is lower than some past linearly polarized fractions detected in BHXBs \citep[e.g.,][]{1992ApJ...400..304H, 2007MNRAS.378.1111B,2014MNRAS.437.3265C,2015MNRAS.451.3975C}. 

The S/N of all linear polarization intensities, are ${\gtrsim}\,$5 with the strongest detections reaching $P_{\lambda,0}/\sigma_{QU} \sim 25$ (see Table~\ref{tab:fracpol_params}). At all $f_\lambda$, imperfect leakage calibration systematically increases the observed linear polarization fraction. This effect is not included in the calculations of the S/N of linear polarization intensities or the errors on $f_\lambda$ that we present.  While our higher $f_\lambda$ values may be (slightly) overestimated due to imperfect leakage corrections, the lower values could be due to spurious signals and are actually consistent with no linear polarization. Following \citet[][]{2017AJ....154...54H}, the predicted level of spurious linear polarization fraction is Rayleigh distributed with a mean given by
\begin{align}
    f_\text{spur,mean} \approx \sqrt{\frac{\pi}{4N_a}\left(f_\text{true}^2 + N_a\left[(S/N)_I\right]^{-2}\right)},
\end{align}
where, $N_a$ is the number of antenna in each sub-array ($N_a$=11/13 for Sub-array 1/2, respectively), $(S/N)_I$ is the Stokes $I$ signal-to-noise ratio of the leakage calibrator at the frequency of interest (with a 16\MHz leakage solution bandwidth), and $f_\text{true}$ is the true linear polarization fraction of the leakage calibrator. For $f_\text{true}$, we adopted the mean linear polarization fraction from the VLA polarization calibrator catalog\footnote{The VLA catalog can be found here; \url{http://www.vla.nrao.edu/astro/evlapolcal/index.html}}, corresponding to, $0.04\%$, and $0.17\%$ for the 5/7 and 21/26\GHz bands respectively. We measured the Stokes $I$ signal from an image plane analysis using \texttt{imfit}. Due to the leakage calibrator’s large Stokes $I$ flux densities and our sparse $uv$-coverage (from a single scan, 13-element sub-array), the Stokes $I$ images are dynamic range limited. As a result, we chose to use $\sigma_{QU}$ as the noise value in $(S/N)_I$, as opposed to $\sigma_I$. The Stokes $Q$ and $U$ images were not dynamic range limited, and thus would better quantify the instrumental noise that also affects the Stokes $I$ data. We present the spurious linear polarization parameters in Table \ref{tab:spur_params}. The differences between the two sub-arrays are the result of elevated noise in sub-array 2. We note that the minimum linear polarization fraction we detect in each base-band is approximately equal to the predicted values of $f_\text{spur,mean}$.

\begin{figure*}
    \centering
    \includegraphics[width=0.95\linewidth]{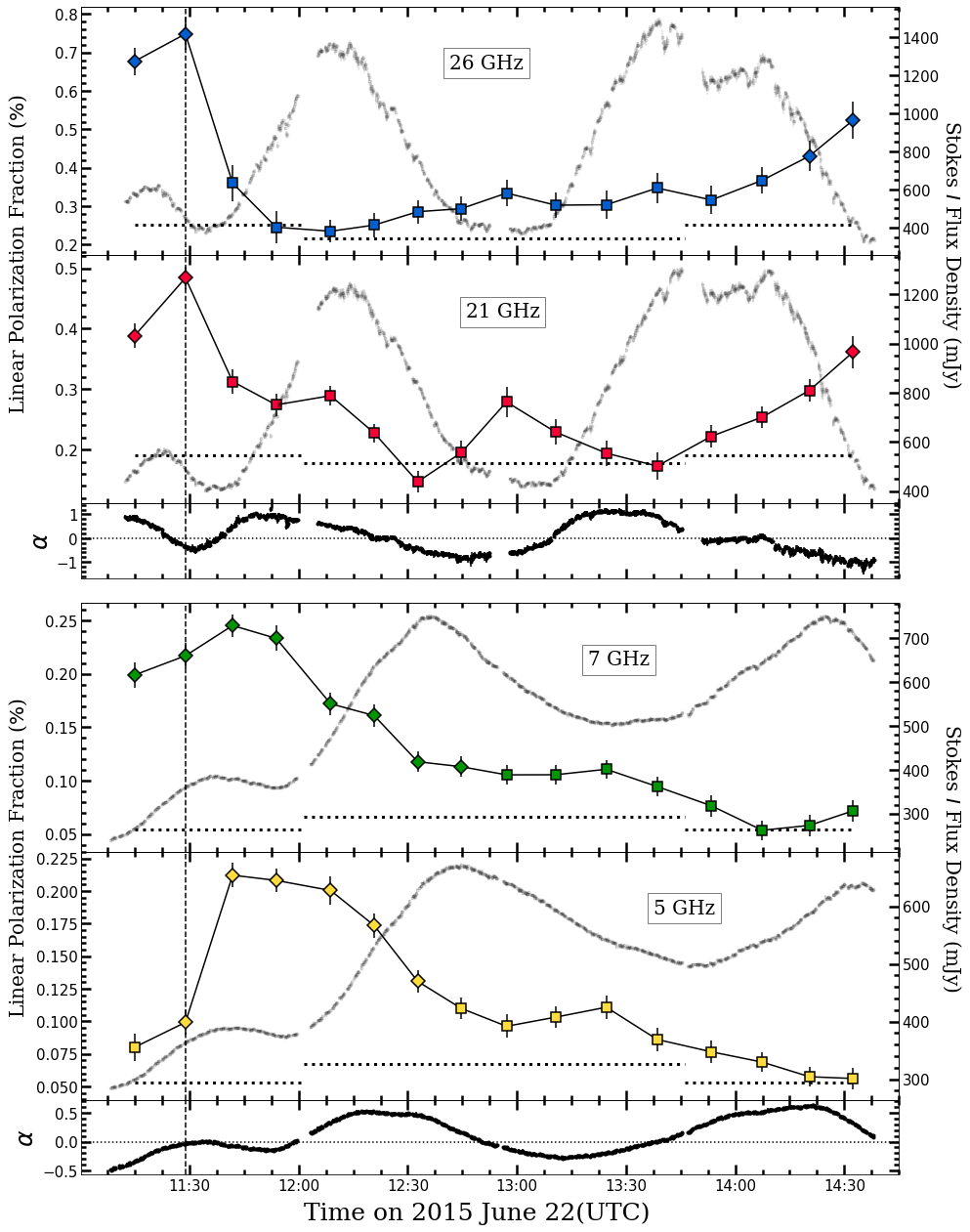}
    \caption{Temporal evolution of the linear polarization fraction; 26\GHz (top), 21\GHz (2nd from the top), 7\GHz (4th from top), 5\GHz (5th from top). The two-point spectral indexes of the 21/26\GHz (3rd from top) and the 5/7\GHz (bottom), show the simultaneous evolution of the absorption conditions. The vertical dashed lines across all panels  highlight the time of the maximum fractional polarization in the 26\GHz base-band. The horizontal dashed lines highlight the value of the mean spurious linear polarization fraction for each base-band; the discontinuities are the result of elevated noise in Sub-array 2. The diamond markers correspond to SL$\geq90\%$ , and the squares to SL$<90\%$. The grey curves display the simultaneous Stokes I flux density evolution for each base-band. We can see that the linear polarization fraction exhibits a similar frequency-dependent delay as the Stokes $I$ light curves, and is offset from the (Stokes $I$) maxima.}
    \label{fig:frac_pol}
\end{figure*}

Henceforth, we define the \textit{significance level} (SL) as the probability that a detection is not the result of a purely spurious signal (See Fig.~\ref{fig:frac_pol}, horizontal-dotted lines). The significance levels for all maxima are $>99\%$, confirming that we have observed an intrinsic polarized signal in each base-band; the SLs for each time bin are tabulated in Table~\ref{tab:fracpol_params}). Only one scan per sub-array was used to correct leakage, and leakage converts Stokes $I$ into $P$. This leads to a single offset in fractional linear polarization in the absence of noise. On the other hand, imperfect leakage can potentially lead to dynamic systematic-error-induced changes in the measured EVPA due to parallactic rotation. 

\begin{table}
    \centering
    \caption{Table of spurious linear polarization properties. All symbols adopt their definitions as defined in the text.}
    \label{tab:spur_params}    
    \begin{tabular}{ccccc}
    \Xhline{3\arrayrulewidth}
         Sub-array & $N_a$ & Base-band & $(S/N)_I$  & $f_\text{spur,mean}$ $(\%)$ \\
    \Xhline{3\arrayrulewidth}
    \multirow[t]{3}{*}{1} & \multirow[t]{3}{*}{13} & \phn5\GHz & 1702 & 0.053\\
    & & \phn7\GHz & 1656 & 0.054  \\
    & & 21\GHz & \phn516 & 0.18\phn  \\
    & & 26\GHz & \phn418 & 0.22\phn  \\
    \hline
    \multirow[t]{3}{*}{2} & \multirow[t]{3}{*}{11} & \phn5\GHz & 1331 & 0.067\\
    & & \phn7\GHz & 1335 & 0.066  \\
    & & 21\GHz & \phn480 & 0.19\phn  \\
    & & 26\GHz & \phn357 & 0.25\phn  \\
    \Xhline{3\arrayrulewidth}
    \end{tabular}
\end{table}

\subsection{Rotation Measure and EVPA}
\label{sec:rm_pol}
\begin{figure*}
    \centering
    \includegraphics[width=0.885\linewidth]{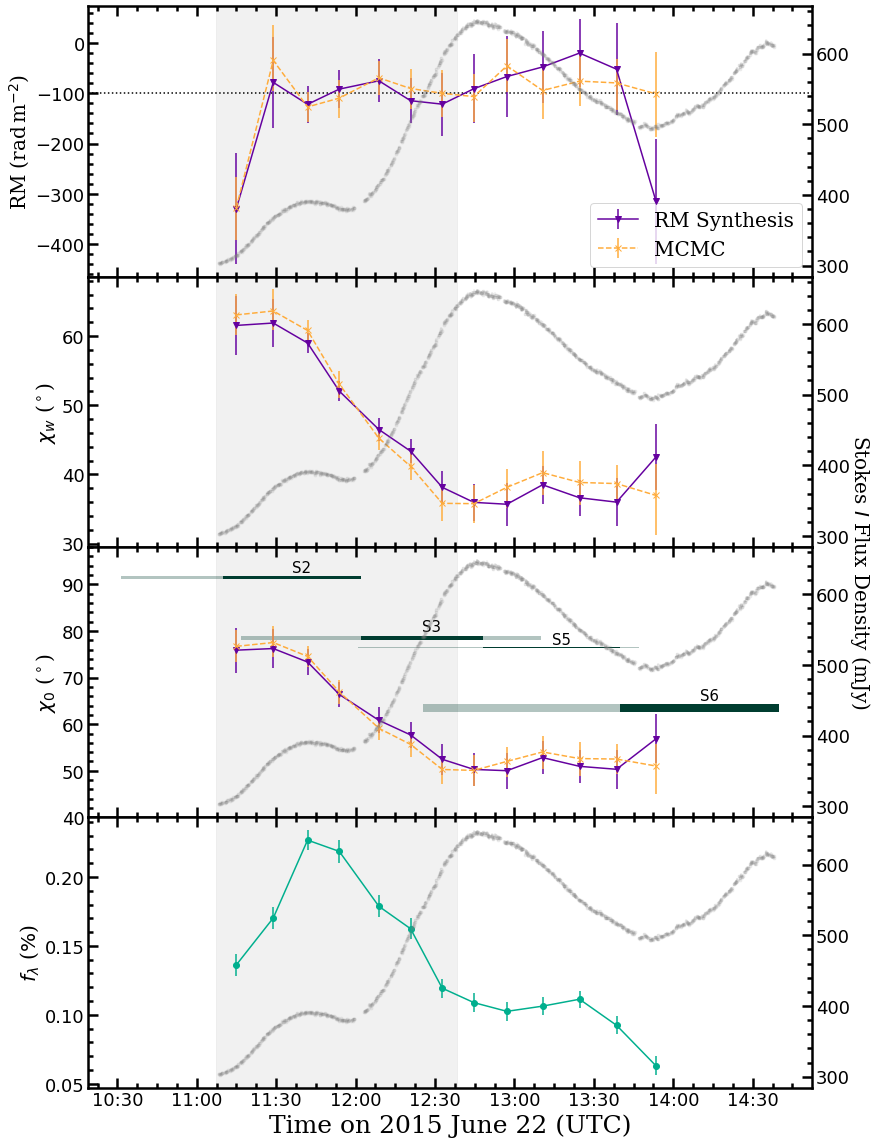}
    \caption{Polarization properties measured from the 5 and 7$\,$GHz base-band observations, for both the MCMC and RM synthesis routines. The vertical shaded region corresponds to the detections with an average significance level ${\geq}\,90{\%}$ between the 5 and 7 GHz base-bands. The underlying grey curve in each panel is the average Stokes I light curve between the 5 and 7$\,$GHz base-bands. There is strong agreement between the two polarimetric methods. (top) The rotation measure; the horizontal dotted line shows the weighted average of the rotation measures (${\sim} -100\radPerSqm$). (2nd from top) The observed EVPA de-rotated to the weighted mean of all $\lambda^2$ channels. (3rd from top) The intrinsic EVPA. The horizontal bars show the PAs ($+90\arcdeg$) of the dominant VLBA ejecta identified in \citet{2019Natur.569..374M}. The length of the bar span the times between the ejection time and when an ejected component is not longer detected, and the darker part of the bar shows when it was the brightest ejected component (excluding the compact core; see Figure \ref{fig:VLBA}). The vertical size of the bars is fixed at the uncertainty in the PA. We adopt the naming conventions from the original paper. (bottom) The average linear polarization fraction between the 5 and 7$\,$GHz base-bands. We can see the rotation measure is constant and the EVPA exhibits a $\sim30\arcdeg$ rotation between $\sim$11:30 and 12:30 during the decay of the polarization fraction maximum.}
    \label{fig:rm_pol}
\end{figure*}

The derived rotation measures exhibit stochastic variability, with values between $-330 \lesssim {\rm RM} \lesssim -20\radPerSqm$ (top panel, Fig.~\ref{fig:rm_pol}), and show a strong agreement between the two polarimetric methods in almost all time bins (the only $\gtrsim1\sigma$ disagreement occurs in the final, lowest S/N time bin). The weighted means of the rotation measure, are $-100\pm16\radPerSqm$ for the RM synthesis method and $-100\pm12\radPerSqm$  for the MCMC method. We applied a simple variability analysis by calculating the $\chi^2$ statistic against a constant RM model equal to the weighted mean. The $\chi^2$ values of 10.6 (RM Synthesis) and 17.4 (MCMC) for 12 degrees of freedom are consistent with a constant rotation measure at probabilities of $56\%$ and $13\%$, respectively.

When we use the default RM synthesis prescription (i.e., quantifying the noise in the dispersion function with an appropriately scaled median absolute deviation rather than the rms), the detection uncertainties reduce by a factor of ${\sim} 2$, and the RM Synthesis $\chi^2$ value become significantly larger than the MCMC value (53.5/12), consistent with a variable rotation measure. However, this choice also increases the population of ${>} 5\sigma_\text{RM}$ components to ${\gtrsim}10$ for each FDF, with rotation measure magnitudes  ${\sim} 10^3\,{-}\,10^5\radPerSqm$. These rotation measures are characteristic of extremely particle-rich lines of sight (e.g., towards the Galactic centre) and, historically, have not been observed in outbursting BHXBs. Our phase calibrator, a source with $\sim 2\%$ linear polarization, showed a similar population of secondary components. We find it very unlikely that these sources would exist while evading detection during recent Galactic rotation measure analyses \citep{2012A&A...542A..93O,2015A&A...575A.118O,2020A&A...633A.150H}. Therefore, we propose that these components are artifacts from imperfect $\lambda^2$ sampling (cf., the effects of poor/patchy $uv$-coverage during synthesis imaging; \citealt{taylor1999synthesis}) or a systematic effect in the modern RM synthesis routine(s). We conclude our decision to use the FDF rms noise is more reflective of the statistical significance of each detection, and that we cannot identify any significant rotation measure variability from V404 Cyg. As a result, for our analyses, we have adopted a constant rotation measure equal to the (inverse-variance) weighted mean of rotation measures across all time bins; i.e., $\text{RM} = -100\pm 16 \, (12) \radPerSqm$ for the RM Synthesis (MCMC) method. 

Both the observed EVPA ($\chi_w$; Second panel, Fig.~\ref{fig:rm_pol}) and intrinsic EVPA ($\chi_0$; Third panel, Fig.~\ref{fig:rm_pol})  exhibit a clear temporal evolution, with strong agreement between the RM synthesis and MCMC routines. Moreover, due to the stable rotation measure, this evolution suggests an intrinsic change in the (polarized) emission environment. In the SL$\,{\geq}\,90\%$ regime, both observed and intrinsic EVPA evolve gradually, with a ${\sim}\,30\arcdeg$ change.
The intrinsic EVPA evolves from  ${\sim}\,80\arcdeg$ to ${\sim}\,50\arcdeg$ between 11:30 and 12:30 UTC. The intrinsic EVPA then stabilized at the ${\sim}\,50\arcdeg$ for the remaining time bins.  

\begin{figure*}
    \centering
    \includegraphics[width=0.85\linewidth]{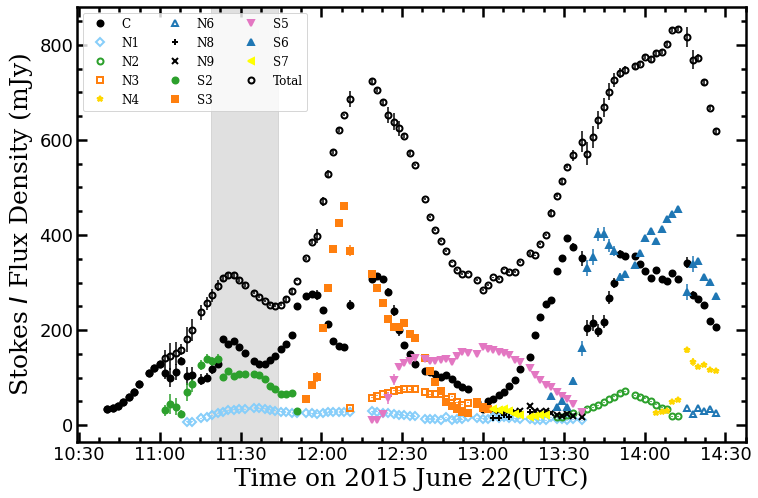}
    \caption{$15.6\,$GHz VLBA light curve using the data from \citet{2019Natur.569..374M}. The open black circles represent the total integrated the flux density, the solid black circles represent the core flux density, and all other marker types represent a single spatially-resolved component. We adopt the naming convention from the original paper. The grey shaded region corresponds to the three time bins that encompass the $7\,$GHz fractional polarization peak; we shifted the region back in time by $10\,$minutes to account for the delay between the bands (see Appendix ~\ref{sec:appendix_delays}). We can see that at any point in time the total integrated (VLBA) flux density is a superposition of multiple radio-bright (and potentially polarized) components that are unresolved in our VLA observations.}
    \label{fig:VLBA}
\end{figure*}

\section{Discussion}
\label{sec:discussion}
In this section we describe the short timescale evolution of our polarimetric results and compare the observed behaviours to the 1989 outburst. Moreover, we correlate this evolution with total intensity light curves, high (spatial) resolution imaging, and optical polarization detections. When analysing the connection between the polarization flaring and the high resolution radio imaging, we limit our discussion to the spatially-resolved VLBA components that dominate the VLBA light curve at a given time. This implicitly assumes that any resolved polarized flux density would track the resolved Stokes $I$ flux density (i.e., $P_{0,\lambda} \propto I_\lambda$). Although we make this assumption, we cannot rule out the possibility that the dominant VLBA components are unpolarized and the sub-dominant components (with average Stokes I flux densities $\lesssim10\%$ of the dominant counterparts) are the source of the polarized emission. 

\subsection{Linear Polarization Fraction}
\label{sec:f_pol_discussion}
The majority of BHXB outbursts have measured linear polarization fractions of ${\gtrsim} 2\%$ at $1\mbox{--}10\,$GHz  \citep[e.g.,][]{2003Ap&SS.288...79F,2007MNRAS.378.1111B,2015MNRAS.451.3975C}, with rare cases reaching appreciable fractions of the theoretical limits \cite[e.g., the ${\sim} 50\%$ detections of XTE J1752$-$223 and Swift J1745$-$26;][]{2013MNRAS.432..931B,2014MNRAS.437.3265C}. Even when considering the typical reduction compared to the theoretical maxima, our measured maximum linear polarization fraction (${\sim} 0.2\%$ in the 5/7\GHz base-bands) for V404 Cyg is a factor of ${\sim} 10$ less than a standard, weakly polarized signal during a BHXB outburst. However, we acknowledge that past outbursts with comparably weak polarization fractions may not have had sufficient S/N for a clear detection and/or the larger (average) polarization fractions may suffer from a publication bias where strongly polarized outbursts are more often introduced within the literature.

Observers caught a glimpse of a comparably low polarization fraction during the monitoring of the 1989 outburst of V404 Cyg. The first day of polarization observations --- 1989 June 1, during the decay of the ``major synchrotron bubble event'' (i.e., the ejection of a bright cloud of synchrotron emitting plasma) --- recorded the lowest polarization fraction of the entire campaign, measuring $0.4\pm0.1\%$ at central frequencies of 4.9 and 8.4\GHz. During the decay of the 1989 outburst, the radio emission exhibited an inverted spectrum and linear polarization fraction of ${\sim} 3\%$, characteristic of a typical compact jet \citep[][]{1992ApJ...400..304H}. The polarized signal was consistently detected for $50\dys$ (between 1989 June 1 and 1989 July 18) before the flux density decayed below the detection threshold. In contrast, we did not observe an increase to a few percent polarization fraction during the decay of the 2015 outburst. The 2015 July 2 epoch places a 99\% confidence interval upper limit of ${\sim} 1\%$ on the 5/7\GHz polarization fraction, consistent with the 2015 June 22 observations, but below the level seen in the compact jet during the 1989 outburst. Furthermore, none of the epochs in \citet{2017ApJ...834..104P} showed any polarized emission, although we note that the upper limits are significantly larger than the maximum polarization fraction detected during the 1989 outburst. 

The Stokes $I$ flux density of V404 Cyg decayed significantly faster in the most recent outburst, taking ${\sim} 30\dys$ in 2015, as opposed to ${\sim} 300\dys$ in 1989. \citet{2018MNRAS.475..448T} suggested that the more rapid decay was the result of the strong winds originating from the accretion disk \citep[detected by][]{2016Natur.534...75M} rapidly depleting the disk and leaving less matter to fuel the jets. Other factors may have also played a role; these could include the total mass reservoir built during the quiescent periods prior to the two outbursts --- 33/26 years for the 1989/2015 outbursts respectively --- or differences in the total mass accreted during the bright outburst phases. In contrast, the polarization fraction depends on the structure of the jet(s), and is only indirectly related to the Stokes $I$ flux density (i.e., ideal, optically-thick synchrotron emission at 1\Jy vs.\ 1\mJy would both have a linear polarization fraction of 10\%). Therefore, the ${<}1\%$ upper limits on 2015 July 2 when the radio emission was dominated by a compact jet (compared to ${\sim} 3\%$ during similar epochs of compact jet dominance in 1989), suggests the most recent outburst had a less-ordered magnetic field in the jet or suffered from higher depolarization due to independent unresolved components within the VLA beam.

There are two clear features of the linear polarization fraction: (i) it is continuously weak (${<}1\%$) regardless of the time bin; (ii) it evolves in time, with maxima and minima linear polarization fractions (in each base-band) separated by a factor of ${\sim} 5$. 

\subsubsection{Origin of Low Linear Polarization Fraction}

The short times between flares, the precession of the jet axis, and the energetics required for such a luminous outburst are characteristic of a complex (magnetic and geometric) environment, and are expected to inhibit strongly polarized emission. At a spatial resolution of ${\sim} 1\AU$ the core emission identified by the VLBA could arise from  an unresolved population of ejecta (likely on top of a compact jet). The jet axis precession would cause these ejecta to have variable PAs, and, assuming similar internal magnetic fields, variable EVPAs. The superposition of the unresolved (and resolved) ejecta in our VLA observations will decrease the polarization fraction, unless all unresolved components have the same polarization fraction and EVPA. The effects of multi-component superposition (i.e., when the coherence length of the magnetic field is significantly smaller than the angular resolution) was seen in the recent Event Horizon Telescope (EHT) observations of M87; the lower spatial resolution of ALMA reduced multiple components with linear polarizations of ${\gtrsim} 20\%$ (resolved with the Event Horizon telescope) to a net polarization fraction of ${\sim} 2\%$ for the M87 core \citep{2021ApJ...910L..12E}. 

In the 2015 outburst of V404 Cyg, the maximum linear polarization fraction in each base-band decreases as the frequency decreases. Here we consider the maximum of each frequency because of the potential time delays between base-bands. A decreasing polarization fraction with decreasing frequency is a common characteristic of Faraday depolarization. In particular, sources with strong Faraday rotation within their emission regions can appear depolarized (in Section ~\ref{sec:rm_discussion} we find that the jet itself may be a strong source of Faraday rotation). Faraday depolarization has been well established in radio studies of AGN \citep[e.g.,][]{2018A&A...613A..74P} and was observed for the candidate BHXB SS 433 \citep{2004MNRAS.354.1239S}. While the complexity of the spectral and temporal evolution of the 2015 outburst of V404 Cyg makes it difficult to determine if we are in fact seeing Faraday depolarization, here we make some simple calculations. The simplest Faraday screen geometry \citep[i.e., a single uniform slab of synchrotron emitting plasma][]{1966MNRAS.133...67B,1998MNRAS.299..189S} predicts a depolarization of $\Delta f_\lambda/f_\lambda \sim 10\%$ between the 26 and 5\GHz base-bands. Therefore, a more complex model \citep[see, ][]{2018A&A...613A..74P} would be required for Faraday depolarization to explain the $\sim70\%$ ($0.75\%$ at 26\GHz to $0.22\%$ at 5\GHz) depolarization we have observed (such an analysis is beyond the scope of this paper). 

We cannot ignore the possibility that during this outburst, the magnetic fields in the jet(s) are intrinsically more disordered than typical BHXB outbursts. The BHXB GRO J1655$-$40 entered a multi-flaring highly-luminous state during its 1994 outburst, similar to the 2015 outburst of V404 Cyg (although the decay timescales of each flare were significantly longer in GRO J1655$-$40). However, GRO J1655$-$40 reached a maximum 4.9 and 8.4 \GHz linear polarization fraction of 1--10\% with linearly polarized variability as high as $\Delta f_\lambda \sim 4\%$ on timescales less than half a day, suggesting that weakly polarized emission is not an inherent aspect of multi-flaring outbursts \citep{1995Natur.375..464H,2000ApJ...540..521H}.

\subsubsection{Origin of Temporally Evolving Linear Polarization}
A transition of the absorption conditions (e.g., from optically-thick to optically-thin synchrotron emission) of a dominant polarized component will result in a temporally evolving polarization fraction \citep[e.g., as seen in Swift J1745-26;][]{2014MNRAS.437.3265C}. During these transitions, we expect the intrinsic EVPA to rotate by 90\arcdeg. For optically-thin synchrotron emission, the EVPA and the magnetic field vector are perpendicular \citep[][]{2011hea..book.....L}, and for optically thick synchrotron emission, the EVPA tracks the direction of the magnetic field \citep[see,][and references therein]{1969ARA&A...7..375G}. The EVPA will thus rotate as the source transitions from $\tau\sim10$ to  $\tau\sim0.5$, where $\tau$ is the optical depth; this takes about half the rise timescale of a vdL plasmoid \citep{1970ApJ...161...19A}. We do not observe a ${\sim} 90\arcdeg$ rotation of the intrinsic EVPA, at any time during our monitoring (see Section \ref{sec:evpa_discussion}). Moreover, we know that the light curves are a superposition of multiple short-lived ($\lesssim 1.5$\hrs) ejecta, and a compact core, further reducing the plausibility of a single component origin for each radio flare.

An ensemble of polarized components with evolving optical depths can exhibit a more complex evolution. As an investigation, we calculated the two-point spectral indexes for the 21/26\GHz and 5/7\GHz VLA observations (see bottom panels of Fig.~\ref{fig:frac_pol}). We are unable to disentangle the emission from the multiple unresolved components (seen in the VLBA), and, as a result, we are measuring the ``net'' spectral index. Moreover, we are measuring a simultaneous spectral index, which may be less appropriate for rapidly evolving ejecta. An optically thick ``net'' spectral index ($\alpha > 0$) requires that a sub-population of the unresolved components are optically thick (with the inverse being true for optically thin, $\alpha < 0$, spectral indexes). The spectral indexes show an evolution in time, exhibiting multiple transitions of the absorption conditions, consistent with an ensemble of evolving components, with both optically thick and optically thin sub-populations. Intuitively, one might expect that a negative ``net'' spectral index measured would correspond to a higher contribution of optically thin synchrotron emission, and, as a result, a higher polarization fraction. The peak polarization fraction does in fact coincide with a negative spectral index; i.e., $\alpha\sim-0.2$ and $-0.5$ in the 5/7\GHz and 21/26\GHz base-bands, respectively (Fig ~\ref{fig:frac_pol}). Furthermore, the late time rise seen in the 21/26\GHz base-bands ($\sim$14:00--14:30), also coincides with a negative spectral index ($\alpha\sim-1$). However, at $\sim$12:45 and 13:15 in the 5/7\GHz and 21/26\GHz base-bands, we also have $\alpha\sim-0.3$ and $-1$. During these times the polarization fraction shows a (weak) peak at 21\GHz, with marginal features at 5/7\GHz, and no evolution at 26\GHz (i.e., a ``missing'' polarization peak). Therefore, we are unable to conclusively connect the spectral index to the polarization fraction evolution.

Comparing the short time-scale temporal evolution to the 15.6\GHz VLBA light curves (reproducing data from \citealt{2019Natur.569..374M} as Fig.~\ref{fig:VLBA} of this paper), we do not see any clear connection between the resolved components and the evolution of the polarization fraction, and cannot distinguish between a polarized core, polarized ejecta, or a combination of the two. However, the ``missing'' polarization peak coincides with a period of time when the S5 component clearly dominates the VLBA light curve. It is possible that S5 was less polarized than the components that preceded and followed its ejection. As a result, a complete explanation of the polarization fraction evolution may require a combination of evolving optical depths, and intrinsic differences between the different polarized components launched at different times. Regarding a potential intrinsic evolution, \citet{2007MNRAS.378.1111B} expanded upon the shock-in-jet picture outlined in \citet{2004MNRAS.355.1105F}, suggesting that the collisions between ejecta temporarily disorder the magnetic field lines while producing shock fronts that propagate through the ejecta, reestablishing a dominant field direction at later times. \cite{2016MNRAS.463.1822S} proposed a similar mechanism to explain the behaviour of the polarized optical emission during V404 Cyg's 2015 outburst. A flare in the optical polarization fraction that preceded a 16$\,$GHz radio flare, was attributed to the compression of the jet's magnetic field by many small shocks travelling along the jet axis. The existence of these shocks is consistent with the detection of sub-second optical flares by \citet{2016MNRAS.459..554G} during the same time period. The polarization flare was attributed to ``a major ejection event'' that followed optical flaring that began a couple of hours earlier; i.e., a large outflow imprinted with the recently ordered magnetic field.

In both of the scenarios proposed by \cite{2007MNRAS.378.1111B} and \cite{2016MNRAS.463.1822S}, the ordering of the magnetic field is a result of multiple colliding components, and, as a result, the timescales separating collisions would have to be significantly shorter than the precession period of the jet. This is a plausible theory if the sub-second optical flaring is characteristic of the collision timescales. Any such model would also need to explain the temporal offset between the Stokes $I$ and polarization fraction peaks.

\subsection{Intrinsic EVPA}
\label{sec:evpa_discussion}
In the first few days of the 1989 outburst, the EVPA evolved through a ${\sim} 90\arcdeg$ rotation at 4.9 and 8.4\GHz. This rotation coincided with the transition from an optically-thin to optically-thick radio spectrum. During the 2015 outburst the dominant feature of the intrinsic EVPA evolution is a ${\sim} 30\arcdeg$ rotation that occurs alongside the decay of the maximum polarization (bottom two panels, Fig.~\ref{fig:rm_pol}). This rotation occurs across 6 time bins ($80\mins$) suggesting that a full $90\arcdeg$ rotation would take ${\sim}4\hrs$, a timescale longer than the lifetimes of any of the dominant VLBA components (see Fig.~\ref{fig:VLBA}). Under the assumption that the contemporaneous peak in polarization fraction and rotation of the EVPA arise from a shared mechanism, neither arise from a transition in the absorption conditions of a single component, as was likely observed in 1989.

The precession of the jet axis provides a natural mechanism to explain the rotation of the EVPA. We investigated this possibility by using the change in the position angles of the dominant ejecta as a proxy for the precession of the jet axis. The position angles also exhibited a ${\sim} 30\arcdeg$ rotation, albeit over a longer, $\sim$2$\hrs$, timescale. The ${\sim}30\arcdeg$ rotation begins when the S2 component (PA$\,{\sim}\,1\arcdeg$) is the dominant jet ejection observed by the VLBA (although the core emission is brighter, see Fig.~\ref{fig:VLBA}). In the simplest geometries, compression shocks or velocity-shearing establish dominant field directions parallel or perpendicular to the jet flow's direction of motion \citep[i.e., the PA;][]{1980MNRAS.193..427L,2007AJ....134..799J}. The approximate orthogonality (offset by ${\sim}\,10\arcdeg$) between the initial intrinsic EVPA and the PA during the decay of the S2 component is consistent with optically-thick (optically-thin) synchrotron emission from a magnetic field established by compression shocks (velocity shearing).

The rotation coincides with the emergence of a new, dominant VLBA component (S3) at a position angle of ${-}11.5\arcdeg$. If the rotation from an intrinsic EVPA of $~80\arcdeg$ to $~50\arcdeg$ results from the S2-to-S3 transition, the larger obliquity (${\sim}\,30\arcdeg$) between the intrinsic EVPA and the PA of S3 requires a more complex magnetic field origin \citep[e.g., remnants of helical fields; ][]{2008ApJ...681L..69G}. Following S3 dominance, S5 becomes the dominant ejection, while maintaining a similar PA of ${\sim}\,13.5\arcdeg$. The similarity between the PAs of S3 and S5 is consistent with the stability of the intrinsic EVPA at ${\sim}50\arcdeg$ between 12:45 and 13:45 UTC, assuming similar intrinsic properties. Alternatively, the S6 component has a smaller obliquity when compared to the late time EVPA (${\sim}\,10\arcdeg$), and may be a better measure of the jet orientation, at later times. However, since our observations are the superposition of multiple overlapping components (including a bright, unresolved compact core), there may be, in fact, no relationship between the position angles of the resolved components and the intrinsic EVPAs.

Variability in the EVPA without any change in the jet axis PA (i.e., \textit{rotator events}) has been observed in many AGN \citep[see][and references therein]{1988ARA&A..26...93S}, and a couple of BHXBs \citep[e.g., GRS 1915+105;][]{2002MNRAS.336...39F}. These events are thought to be the result of complex field geometries \citep[e.g., helical magnetic fields;][]{2001ApJ...561L.161G} or internal shocks \citep[e.g.,][]{2008ApJ...681L..69G} producing time-varying magnetic fields. Moreover, complex shock fronts (e.g., conical shock waves) can produce magnetic field orientations that are neither perpendicular nor parallel to the jet axis \citep[see][and references therein]{2007AJ....134..799J}. 

Since the VLBA data did not acquire full polarimetric calibrations, there is no spatially resolved polarimetry that explicitly localizes the dominant polarized component. In the absence of such detections and given the multiple scenarios suggested above, we can neither definitively make connections between the VLBA/VLA observations and the linear polarization properties, nor identify if the polarized emission originates from an ejected component, a compact steady jet, or a time-variable combination of the two. This limits the strength of our claims towards the origin of the polarization and its connection to the evolution of the Stokes $I$ flux density. We note that due to the reduced sensitivity of spatially-resolved data, without a significant increase in the polarization fraction (as was observed in M87), the VLBA would be unable to detect comparably low polarization fractions, even after including the necessary calibrators. 

\subsection{Rotation Measure}
\label{sec:rm_discussion}

\begin{figure}
    \centering
    \includegraphics[width=0.85\linewidth]{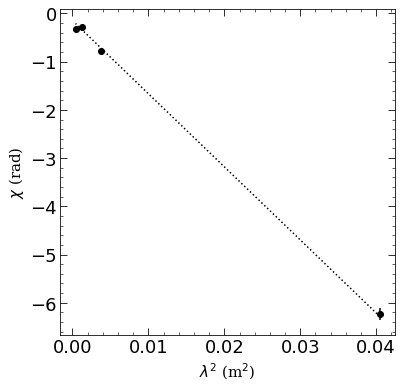}
    \caption{Linear fit to the observed EVPAs during V404 Cyg's 1989 outburst; the data was adapted from \citealt{1992ApJ...400..304H}. To account for the wrapping of the EVPA at large values of $\lambda^2$ we applied a $-2\pi$ correction to the 1.49$\,$GHz observation. We used \texttt{scipy.optimize.curve\textunderscore fit} for our linear fit.}
    \label{fig:HH92}
\end{figure}

The rotation measure quantifies the amount of Faraday rotation affecting a linearly polarized emission signal, and is related to the internal properties of the plasma along the line of sight (i.e., its Faraday screens). The RM is related to the electron number density, $n_e$, the magnetic field oriented parallel to the line of sight (from the source to the observer), $B_{||}$, and the path length $l$. Explicitly, the rotation measure is described by the path integral,
\begin{align}
    \text{RM} = \left[812\int_\text{source}^\text{observer} n_eB_{||}\text{d}l\right]\, {\rm\,rad\,m^{-2}},
\end{align}
where $n_e$, $B_{||}$, and d$l$ are in units of cm$^{-3}$, $\mu$G, and kpc, respectively. The sign of the rotation measure depends on the orientation of the magnetic field; i.e., when the field lines are parallel (anti-parallel) to the direction of emission propagation, the sign is positive (negative). Moreover, for Galactic sources, the total rotation measure can have significant contributions from both the diffuse interstellar medium (ISM) and the local environment. Detecting a large local component necessarily implies a high density, or strongly magnetic environment to account for the reduced path length when compared to the ISM.

During the previous outburst in 1989, \citet{1992ApJ...400..304H} measured a constant observed EVPA in four frequency bands for the majority of the (${\sim} 50\dys$) polarization monitoring. The weighted averages from these observations were; $3 \pm 7\arcdeg$, $-44 \pm 1\arcdeg$, $-16 \pm 1\arcdeg$, and $-18\pm2\arcdeg$, at central frequencies of 1.49, 4.9, 8.4, and 14.9\GHz, respectively. The observed EVPAs are linear with respect to $\lambda^2$ (see Fig.~\ref{fig:HH92}), with a slope (i.e., rotation measure) of $-151\pm11\radPerSqm$. During the first ${\sim}2\mbox{--}3 \dys$ of polarization detections, the EVPAs at 4.9 and 8.4\GHz  exhibited a $90\arcdeg$ rotation. The two-point slope of these angles ($39\pm4\arcdeg$ and $-60\pm6\arcdeg$ at 4.9 and 8.49\GHz, respectively) shows a consistent rotation measure of $-150\pm50\radPerSqm$. The matching rotation measures, even with a changing EVPA, implies a constant Faraday screen. The magnitude, orientation, and stability of the 1989 rotation measure is similar to our RM measurement during the 2015 outburst ($-100 \pm15\radPerSqm$); however, the former is detected over much longer timescales. The rotation measures during the 1989 and 2015 outbursts are marginally consistent (at the $\sim2.7\sigma$ level). As a result, we are unable to conclusively identify temporal variability of the rotation measure (e.g., as was seen in the 1994 outburst of GRO J1655$-$40; \citealt{2000ApJ...540..521H}). Had we identified temporal variability, we could rule out the scenario that both outbursts are behind a constant, purely Galactic Faraday screen.

Multiple Galactic RM models predict a negative rotation measure along the line of sight to (and beyond) V404 Cyg; $-30\pm10 \radPerSqm$ \citep{2012A&A...542A..93O}, $-40\pm20 \radPerSqm$ \citep{2015A&A...575A.118O}, and $-130\pm50\,\radPerSqm$ \citep{2020A&A...633A.150H}. Assuming a constant Galactic magnetic field, the Galactic RM would be dominated by distances well beyond the ${\sim} 2.4 \kpc$ distance to V404 Cyg. For both standard models of electron distributions \citep{2003astro.ph..1598C,2017ApJ...835...29Y} the \textsc{PyGEDM} tool \citep{2021PASA...38...38P} indicates much larger dispersion measures (which are proxies for the integrated electron column density) at 10\kpc ($269\pm16\pc\cms^{-3}$) than at 2.4\kpc ($32\pm4\pc\cms^{-3}$).
Thus, our measured value of ${\approx}-100\radPerSqm$ suggests, either an inversion (or multiple inversions) of the Galactic magnetic field along our line of sight, or an intrinsic rotation measure component local to the source. The morphology of the Galactic magnetic field is poorly constrained, with different models predicting radically different structures \citep[see,][]{2015ASSL..407..483H,2019Galax...7...52J}. As an example, the ``zeroth"-order model by \citet{2011ApJ...728...97V} predicts a parallel Galactic magnetic field within the first ${\sim} 4 {\rm \, kpc}$ along the line-of-sight containing V404 Cyg, inverting to an anti-parallel orientation at larger distances and producing the net-negative rotation measure. Conversely, the model by \citet[][]{2012ApJ...757...14J}, predicts two large-scale inversions along the line of sight of interest; an initial anti-parallel magnetic field (and a negative rotation measure at the position of V404 Cyg), an inversion to a parallel orientation at intermediate distances, followed by a second inversion back to an anti-parallel orientation. However, using the standard approximation, $|B_{||,\text{avg}}| = |\text{RM}/(0.81\text{DM})|$, we can estimate a mean parallel magnetic field magnitude of $3.8\pm0.7\muG$, which is larger than the total ($\sqrt{B_{||}^2 + B_{\perp}^2}$) Galactic magnetic field magnitudes predicted by both 
\citet[][$\sim0.1\muG$]{2011ApJ...728...97V} and 
\citet[][$\sim1.0\muG$]{2012ApJ...757...14J}. Given our estimate for the mean parallel magnetic field strength along the line of sight towards V404 Cyg, there exists three physical explanations: (i) the mean electron number density is larger than predicted by the standard dispersion models along this line of sight; (ii) the mean magnetic field strength within the ISM is stronger than predicted by Galactic magnetic field models along this line of sight; or (iii) there is a local Faraday screen that likely resides within the jets themselves. Here we investigate the source of a (potential) rotation measure component local to V404 Cyg.

A local rotation measure component is the result of either a foreground Faraday screen (e.g., created by disk outflows) or a rotation from within the emission regions themselves (e.g., the compact core or jet ejecta). \citet{2016Natur.534...75M} detected a strong, continuous wind originating from the accretion disk. Assuming that the wind creates a foreground Faraday screen with an electron number density that follows an inverse-square scaling, $n_e \equiv n_0(l/l_0)^{-2}$, and a typical ISM magnetic field strength ($B_{||}{\sim}\,2\,\mu$G; \citealt{2015ASSL..407..483H}), equation (5) simplifies to, 
\begin{align}
    {\rm\,RM}=1624 \, n_0 l_0^2\left(\frac{1}{l_0}-\frac{1}{l_\text{max}}\right) {\rm\,rad\,m^{-2}},
\end{align}
where the wind-fed Faraday screen occupies the space between $l_0$ and $l_\text{max}$ along our line of sight. We approximate $l_\text{max}\sim v\Delta t \sim 8\,$AU using the measured wind velocity ($v\sim2000 {\rm\,km\,s^{-1}}$; \citealt{2016Natur.534...75M}) and the time interval between the start of the outburst and our observations ($\Delta t \sim7\,\dys$). We adopt the VLBA angular resolution of 1$\,$AU, as a conservative estimate of $l_0$ for compact core emission. The jet ejections with well constrained inclination angles are S2 ($\sim40^\circ$), S3 ($\sim30^\circ$), and S6 ($\sim15^\circ$); all three ejecta have an angular separation of $\sim0.5\,$milliarcseconds during their flux density peaks \citep{2019Natur.569..374M}. The distance to V404 Cyg is 2.39 kpc, and, therefore, $l_0\sim2-5\,$AU for jet ejections. For a wind-fed Faraday screen to produce our observed rotation measure ($|\text{RM}|\sim100\,\radPerSqm$), we require $n_0\sim(6-15) \times 10^6\,{\rm cm^{-3}}$. Assuming a 50\% ionized, isotropic, pure hydrogen outflow, launched at a distance of $6\times10^5\,{\rm km}$ from the central black hole, the wind mass loss rate would need to be $\Dot{M} \sim (0.4-2) \times 10^{-6} \Msun \yrs^{-1}$. \citet{2016Natur.534...75M} estimated a wind mass loss rate of $>10^{-13} \Msun \yrs^{-1}$, $\sim$7 orders of magnitude smaller then our calculations. The authors left the estimate as a lower limit because the ionization fraction ($\propto \Dot{M}^{-1}$) may be lower then the assumed value of $f_i = 0.5$, and the launching radius ($\propto \Dot{M}$) may be larger then their assumed value of $R = 6 \times 10^5 {\,\rm km}$. A 7 order of magnitude reduction in the ionization fraction would inhibit Faraday rotation, as the outflow would become neutral. Furthermore, a 7 order of magnitude increase in the launching radius corresponds to an distance of $4\times10^4\,$AU, far exceeding the scale of the system. Therefore, without a highly magnetized wind, or an extremely anisotropic wind coupled with a favourable line of sight, disk winds forming a foreground screen cannot be the origin of the observed rotation measure.

For Faraday rotation internal to the emission environment we look at the recent model of the compact jet from MAXI J1820$+$070 \citep[see, ][for a detailed description of the model]{2022ApJ...925..189Z}. The strength of the magnetic field, and the electron number density scale according to the power-law relations, $B = B_0 \xi^{-b}$ and $n_e = n_0 \xi^{-a} \gamma^{-p}$, where $\gamma$ is the lorentz factor of the synchrotron emitting electrons, and $\xi = \frac{z}{z_0} = (\frac{\nu}{\nu_0})^{-q}$, where $z$ is the position along the jet axis. At $z>z_0$, the jet emits synchrotron radiation; this leads to a break in the spectrum from optically thick to optically thin at $\nu_0$.

We adopt the following values used in the original paper: $b=1.1$; $a=2.2$; $p=2$; $q=0.882$; $B_0 \sim 10^{10}\muG$, $z_0 \sim 3 \times 10^{10} \cms$, $\nu_0 \sim 2.3 \times 10^4\GHz$, $n_0 \sim 3 \times 10^{14} \cms^{-3}$, and we adopt a value of $\gamma = 0.5\,{\times}\,(\gamma_\text{min} + \gamma_\text{max})=386.5$. The model predicts $B  \sim  3  \times 10^6 \muG$ and $n_e  \sim 2 \times 10^2 \cms^{-3}$ at $\nu=6\GHz$. Letting, $B_{||} = 0.5B$, and assuming a uniform Faraday screen, we would require a screen thickness of $d l  \sim 3 \times 10^{-10}\kpc$ to account for the rotation measure. To first-order, this is the same as the radius of the conical jet at position $z$, $R = z\sin\theta$, for the best fit opening angle $\theta \sim 1.5\arcdeg$. Considering, that the best fit orbital inclination angle is ${\sim} 65\arcdeg$, it is reasonable to assume the our line of sight looks partially down the jet axis, and, as a result, $dl > R$. Furthermore, The electron number density could be substantially larger than expected from a typical hard state compact jet if the jet entrains material from the disk winds. Entrainment is a known source of internal Faraday depolarization in AGN \citep[e.g.,][]{2022MNRAS.513.4208S}, and jet-wind interactions have been observed in the BHXB candidate SS 443 \citep{2011ApJ...735L...7B}. Although we are unable to rule out that V404 Cyg has a magnetic field oriented perpendicularly to the line of sight, or significantly different jet parameters when compared to MAXI J1820$+$070 (e.g., a weaker magnetic field), to first-order, it is plausible that the jet itself may act as a strong, local Faraday screen. 

\section{Summary and Conclusions}
\label{sec:conclusions}
In this paper we present our analysis of the multi-frequency (5, 7, 21, and 26\GHz), linear polarization radio data of the BHXB V404 Cyg during its 2015 outburst. The majority of our results and interpretations focused on the behaviour during the bright flaring activity on 2015 June 22, however, we also included the upper limits from six observations during the source's return to quiescence. Using two independent polarimetric methods we extracted the fractional polarizations, observed/intrinsic EVPAs, and rotation measures from the 2015 June 22 data. We tracked the evolution of the polarization properties on timescales ${\sim}13\mins$, constituting one of the shortest timescale polarimetric analyses of a BHXB to date.

By comparing our polarimetric results to the VLA Stokes $I$ light curves modelled by \citet{2017MNRAS.469.3141T} and the simultaneous VLBA observations by \citet{2019Natur.569..374M}, we infer the following properties about the polarization evolution of V404 Cyg:
\begin{itemize}
  \setlength\itemsep{1em}
    
    \item V404 Cyg is weakly polarized, with a maximum polarization fraction that increases with frequency; ${\sim}$0.22, 0.25, 0.5, 0.75\% for the 5, 7, 21, and 26\GHz base-bands, respectively. These maxima are significantly smaller than typically observed in outbursting BHXBs, suggestive of a complex local environment or complex internal magnetic field structure. 
    
    \item The time-evolution of the linear polarization fraction shows a frequency-dependent lag, with low frequencies lagging behind their high frequency counterparts. This behaviour is characteristic of an emission origin within dynamic components (e.g., expanding ejecta or propagating shock fronts). 
    
    \item The maximum polarization fraction is offset from the Stokes $I$ flux density maxima. This suggests an offset between the processes that maximize each quantity. A secondary peak in fractional polarization at 21\GHz after the second flare in Stokes $I$ and the increase in polarization fraction towards the end of the 21/26$\,$GHz base-band observations, provide further evidence of a temporal offset.
    
    \item The decay of the (brightest) polarization fraction peak coincides with a rotation of the intrinsic EVPA. We are unable to conclusively determine if the origin of this feature is the result of an internal change within the polarized components, or the emergence (and decay) of polarized components with different magnetic field structures. 
    
    \item The derived rotation measures show stability in time with an average value of ${\sim} {-}100\radPerSqm$. We investigated the potential of a strong local component, and although we found it plausible, we are unable to conclusively rule out a purely Galactic rotation measure.

\end{itemize}

Overall, our results emphasize the complexity of local (magnetic) environments during highly energetic outbursts. Although we are confident that the observed behaviour cannot be ascribed to the simplest interpretation and models (e.g., intrinsic EVPA swings from changes in the absorption conditions of single components), the limitations of our observations inhibited us from making strong claims about the origin of the polarized emission. These limitations emphasize the importance of spatial resolution during polarimetry, which would enable the identification of the primary source of the polarized emission in multi-component outbursts. X-ray polarimetric observations of black hole X-ray binaries \citep[like those recently done for the black hole X-ray binary Cygnus X-1; ][]{2022arXiv220609972K} probe the accretion disk, corona and perhaps some component from the synchrotron tail of a jet. For the radio-brightest X-ray binary outbursts, there is strong potential to combine such observations with radio through sub-mm polarimetric observations to track temporally evolving polarization properties across the electromagnetic spectrum. However, we note that, like the VLA observations performed here that did not have the angular resolution needed to separate polarization properties from different ejecta, one must carefully consider how the polarization properties from different components will average when making interpretations. If the next ${\sim}1\,$Jy scale outburst of a BHXB is observed with adequate spatial resolution, full polarization coverage, and sufficient sensitivity, such observations have the potential to provide invaluable insight into the magnetic fields that drive accretion-powered jet ejections, emission, and evolution. The new generation of interferometers like the next-generation VLA, the next-generation EHT, and the Square Kilometre Array, will combine high sensitivity and good spatial resolution into a single instrument, making spatially resolved polarimetry a realistic goal for future, bright outbursts.

\section*{Acknowledgements}

We extend our sincere thanks to all of the NRAO staff involved in the scheduling and execution of these observations. We offer a special thanks to Frank Schnitzel for sharing his expertise of polarization observation, and calibration, and data reduction with the VLA and CASA. We thank Cameron Van Eck for helpful discussions regarding Galactic magnetic field models and the Canadian Initiative for Radio Astronomy Data Analysis (CIRADA) RM synthesis routine. Finally, we thank the referee for their insightful and helpful comments. 

This research has made use of software provided by CIRADA. CIRADA is funded by a grant from the Canada Foundation for Innovation 2017 Innovation Fund (Project 35999), as well as by the Provinces of Ontario, British Columbia, Alberta, Manitoba, and Quebec, in collaboration with the National Research Council of Canada, the US National Radio Astronomy Observatory and Australia’s Commonwealth Scientific and Industrial Research Organisation. 

AKH and GRS are supported by NSERC Discovery Grants RGPIN-2016-06569 and RGPIN-2021-0400. Support for this work of AJT was provided by NASA through the NASA Hubble Fellowship grant \#HST--HF2--51494.001 awarded by the Space Telescope Science Institute, which is operated by the Association of Universities for Research in Astronomy, Inc., for NASA, under contract NAS5--26555. TDR acknowledges financial contribution from the agreement ASI-INAF n.2017-14-H.0. GEA is the recipient of an Australian Research Council Discovery Early Career Researcher Award (project number DE180100346). SM is thankful for support from an NWO (Dutch Research Council) VICI award, grant Nr. 639.043.513. TMB acknowledges the financial contribution from grant PRIN-INAF 2019 N.15. RS acknowledges support from grant number 12073029 from the National Natural Science Foundation of China (NSFC).

 AKH and GRS respectfully acknowledge that they perform the majority of their research from Treaty 6 territory, a traditional gathering place for diverse Indigenous peoples including the Cree, Blackfoot, Métis, Nakota Sioux, Iroquois, Dene, Ojibway/ Saulteaux/Anishinaabe, Inuit, and many others whose histories, languages, and cultures continue to influence our vibrant community. The authors also wish to recognize and acknowledge the significant cultural role and reverence that the summit of Maunakea has always had within the indigenous Hawaiian community. We are most fortunate to have the opportunity to conduct VLBA observations from this mountain.

\section*{Data Availability}
Data from the VLA are available through the VLA data archive (Project ID 15A--504): \url{https://archive.nrao.edu/archive/advquery.jsp}. We make our flagging and calibration scripts, and the short timescale imaging and analysis, available at https://doi.org/10.11570/22.0002.



\bibliographystyle{mnras}
\bibliography{bibly} 


\ \par
{\itshape
\noindent\EndAffil}



\appendix

\newpage
\section{Stokes \textit{I} Light Curves and Temporal Delays}
\label{sec:appendix_delays}
\begin{figure*}
    \centering
    \includegraphics[width=0.9\linewidth]{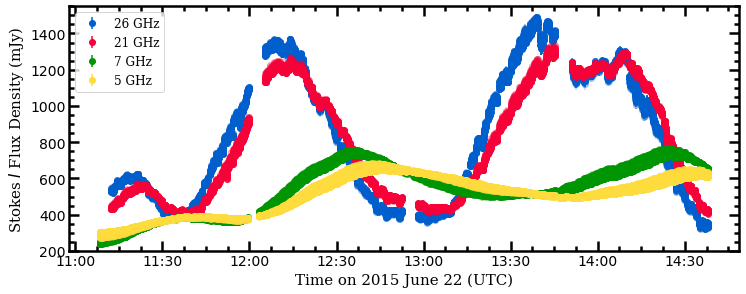}
    \caption{High time resolution Stokes $I$ light curves for our observing base-bands. In each base-band, we have separately plotted the flux densities of the 8 spectral windows. The key features are the three major flares, with the final flare having a twin-peaked structure in the 21/26$\,$GHz base-bands. The lower frequency emission is temporally delayed, peaks at lower flux densities, and has broader flares. These properties are consistent with both vdL ejecta and a shock-in-jet event scenario.}
    \label{fig:light_curves}
\end{figure*}

The Stokes $I$ flux density light curves for all four base-bands share a common ``three-flare'' morphology (see Fig.~\ref{fig:light_curves}), where each flare is composed of multiple unresolved (by the VLA) jet ejecta \citep{2017MNRAS.469.3141T,2019Natur.569..374M}. The temporal delays and longer rise/decay times for the lower frequency flares are consistent with the expected behaviour of expanding vdL bubbles. Considering a single vdl ejection, the flaring results from an evolving optical depth; the peak flux density occurs near the transition from optically thick to optically thin emission, and the peak occurs at earlier times for higher frequencies \citep{1966Natur.211.1131V}. Therefore, broadband observations will mix optically thick (low frequency) and optically thin (high frequency) emission. Optically thick and thin emission are orthogonally polarized, and thus their summation causes a (potentially significant) depolarization effect. To quantify this effect, we measured the delays between flux density peaks (for each of the three flares) from the cross-correlation function (CCF) of the high time resolution (per-spectral window) light curves. Each CCF was generated using the z-transformed discrete correlation function (ZDCF) techniques of \citet{1997ASSL..218..163A}\footnote{FORTRAN code available at \url{http://www.weizmann.ac.il/particle/tal/research-activities/software}.}. We measured delays between flares of ${\sim} 30\text{--}60 \mins$  (${\sim} 2\text{--}4 {\times}\,$ the imaging window) and ${\sim} 7\text{--}15\mins$ (${\sim} 0.5\text{--}1 {\times}\,$ the imaging window) comparing the 5-to-26\GHz and 5-to-7\GHz base-band light curves, respectively. The earliest flare exhibited the smallest delays between bands. For the polarization fraction, the small number of time bins inhibited a similar use of the ZDCF algorithm. However, looking at Figure \ref{fig:frac_pol}, the delays between the polarization fraction peaks are ${\sim} 2$ imaging windows (${\sim} 30 \mins$) and ${\lesssim} 1$ imaging window (${\lesssim} 15 \mins$), for the 5-to-26\GHz and 5-to-7\GHz delays, consistent with the Stokes $I$ behaviour. The 5-to-26\GHz delays are an appreciable fraction of the lifetime of a single ejection ($\lesssim 90\mins$), and, therefore, the full bandwidth (5-to-26\GHz) may have a non-negligible fraction of orthogonal emission, even when considering isolated ejecta.

The situation becomes considerably more complicated when considering the multi-ejecta flares, ejecta collision, and jet precession (as seen in the 2015 outburst of V404 Cyg). Moreover, we note that modelled ejecta exhibit different delays between frequencies (largely due to differences in ejecta expansion velocities). As such, the dominant ejection (in a particular flare) depends on the observing frequency, thereby introducing another frequency-dependent effect on the observed EVPA. While critical, the VLBA observations only provide a snapshot at one frequency. Due to the large temporal separations, we choose to omit the simultaneous linear polarization data from the 21 and 26\GHz base-bands when extracting EVPAs and rotation measures. This is why we only consider the simultaneous linear polarization data from the 5 and 7\GHz base-bands in \S~\ref{sec:rm_pol}, \S~\ref{sec:evpa_discussion}, and \S~\ref{sec:rm_discussion}. We intend to apply temporal corrections to future (broad-band) outbursts with single (or temporally isolated) ejecta as an investigation into the effects these delays have on polarization measurements. 

\section{Calibrator Model}
\label{sec:calibrator_model}

\begin{figure*}
    \centering
    \includegraphics[width=0.95\linewidth]{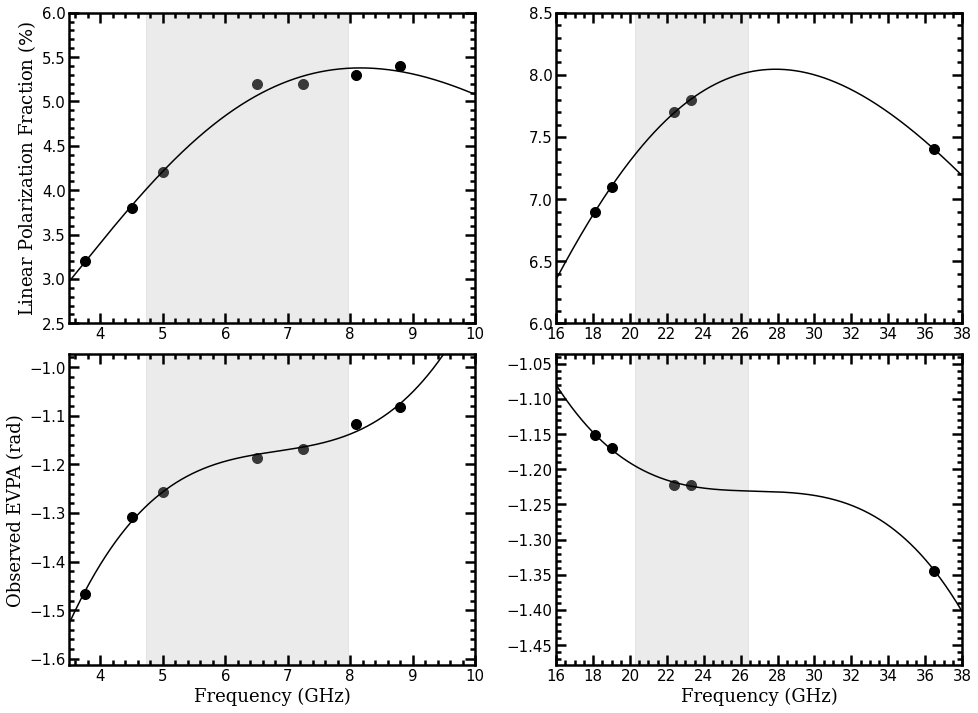}
    \caption{Polarization fraction spectra for 3C48; \textit{left}: 5/7\GHz base-bands and \textit{right}: 21/26\GHz base-bands. The shaded regions highlight the range of frequencies spanned by each observing band. The solid black points are the data from \citet{2013ApJS..206...16P}, and the black line is our polynomial fit. The logarithmic term in equation (3) is exactly equal to the equivalent frequency in GHz.}
    \label{fig:fit_params}
\end{figure*}
For Stokes $I$ calibration, \textsc{casa} includes a repository of spatially resolved model images for many standard calibrators. Our flux calibrator, 3C48, is included in this repository. Each model image describes the flux density distribution at a single, band-dependant, reference frequency (e.g.,  $\nu_\text{ref} = 4.8601\GHz$ for the 4-8\GHz band). During calibration, the model is mapped onto the remaining spectral channels assuming the total flux density follows the flux density scaling relationships of \mbox{\citet{2017ApJS..230....7P}}. It is assumed that the spatial distribution of the relative flux densities remains constant across a band; i.e., for an arbitrary spectral channel with a central frequency $\nu$, the ratio between the total integrated flux, $I_\nu$, and the flux of pixel $i$, $I_{i,\nu}$, is independent of frequency, and thus identical to the ratio at the reference frequency ($I_{i,\nu}/I_\nu \equiv I_{i,\nu_\text{ref}}/I_{\nu_\text{ref}}$). This procedure results in a spatially resolved Stokes $I$ model for every spectral channel.

For our Stokes $Q$ and $U$ calibration, we adopted a similar approach to the default Stokes $I$ prescription. We assumed that the spatial distribution of the linearly polarized flux densities is independent of frequency and that it has the same spatial distribution as the Stokes $I$ repository images. For each spectral channel with a central frequency of $\nu$, we calculated the total Stokes $Q$ and $U$ flux densities according to the following relationships,
\begin{align}
    &Q_\nu = I_\nu f_{\nu}\cos\left(2\chi_{ \nu}\right), \\
    &U_\nu = I_\nu f_{\nu}\sin\left(2\chi_{ \nu}\right),
\end{align}
where $f_\nu$ is the linear polarization fraction, and $\chi_\nu$ is the observed EVPA. We mapped the total flux densities onto each pixel by assuming that $U_{i,\nu}/U_\nu \equiv Q_{i,\nu}/Q_\nu \equiv I_{i,\nu_\text{ref}}/I_{\nu_\text{ref}}$. Our final model consisted of a spatially resolved Stokes $I$, $Q$, and $U$ image (assuming no circular polarization; i.e., Stokes $V=0$) for each spectral channel. We then applied the model to our data using the native \textsc{casa} task \texttt{ft}.

In equation (B1) and (B2), $I_\nu$ was calculated according to \citet{2017ApJS..230....7P}. For $f_\nu$ and $\chi_\nu$, we fit the data presented in \citet{2013ApJS..206...16P}, such that the linear polarization fraction obeys a third-order log-log polynomial, 
\begin{align}
\log(f_\nu) = \sum_{n=0}^3a_n\log\left(\nu_{\rm GHz}\right)^n,
\end{align}
and the observed EVPAs obey a standard third-order polynomial (with frequency),
\begin{align}
\chi_\nu = \sum_{n=0}^3b_n\left(\nu_{\rm GHz}\right)^n;
\end{align}
where $\nu_{\GHz}$ is the observing frequency (in GHz). We decided to fit the polarization fraction with a $3^\text{rd}$-order polynomial in log-space to remain consistent with the $I_\nu$ fits in \citet{2017ApJS..230....7P}, and $\chi$ in linear-space to allow for negative EVPAs. We fit the polarization properties separately for the 5/7\GHz and 21/26\GHz bands, rather than a single fit as performed by \citet{2017ApJS..230....7P}, following extensive discussions with NRAO staff  (F.~Schinzel priv.~comm.). Table \ref{tab:fit_params} contains the third-order polynomial fit for the polarization calibrator model and Fig.~\ref{fig:fit_params} plots the fit over the \citet{2013ApJS..206...16P} observations of 3C48.

\begin{table}
    \centering
    \caption{The third-order polynomial fits for Stokes $I$, the polarization fraction spectra, and the intrinsic EVPA of 3C48. The $I_\nu$ values are taken from \citet{2017ApJS..230....7P}.}
    \label{tab:fit_params}    
    \begin{tabular}{cccccc}
    \Xhline{3\arrayrulewidth}
         Quantity & Band & $a_0$ & $a_1$ & $a_2$ & $a_3$ \\
    \Xhline{3\arrayrulewidth}
         $I_\nu$ & \begin{tabular}{@{}c@{}}\phn\phn5/7\,GHz \\ 21/26\,GHz\end{tabular} & \begin{tabular}{@{}c@{}}$\phantom{-}1.3253$ \\ $\phantom{-}1.3253$\end{tabular}& \begin{tabular}{@{}c@{}}$-0.7553$ \\ $-0.7553$\end{tabular}& \begin{tabular}{@{}c@{}}$-0.1914$ \\ $-0.1914$\end{tabular} & \begin{tabular}{@{}c@{}}$\phantom{-}0.0498$ \\ $\phantom{-}0.0498$\end{tabular} \\
    \hline
          $f_\nu$ & \begin{tabular}{@{}c@{}}\phn\phn5/7\,GHz \\ 21/26\,GHz\end{tabular} & \begin{tabular}{@{}c@{}}$-1.5775$ \\$\phantom{-}1.5927$\end{tabular}& \begin{tabular}{@{}c@{}}$-1.6671$ \\ $-8.9992$\end{tabular}& \begin{tabular}{@{}c@{}}$\phantom{-}4.7686$ \\ $\phantom{-}8.5932$\end{tabular} & \begin{tabular}{@{}c@{}}$-2.8181$  \\ $-2.5275$\end{tabular}  \\
    \Xhline{3\arrayrulewidth}
         Quantity & Band & $b_0$ & $b_1$ & $b_2$ & $b_3$ \\
    \Xhline{3\arrayrulewidth}

          $\chi$ & \begin{tabular}{@{}c@{}}\phn\phn5/7\,GHz \\ 21/26\,GHz\end{tabular} & \begin{tabular}{@{}c@{}}$-3.7987$ \\$\phantom{-}0.9939$\end{tabular}& \begin{tabular}{@{}c@{}}$\phantom{-}1.1141$ \\ $-0.2480$\end{tabular}& \begin{tabular}{@{}c@{}}$-0.1602$ \\ $\phantom{-}0.0092$\end{tabular} & \begin{tabular}{@{}c@{}}$\phantom{-}0.0078$  \\ $-0.0001$\end{tabular}  \\
    \Xhline{3\arrayrulewidth}
    \end{tabular}
\end{table}

\renewcommand{\arraystretch}{1.5}

\section{Phase Calibrator Polarimetric Evolution}
\label{sec:phase_cal_evolution}
Given the low linear polarization fractions we detected in our V404 Cyg observations, we checked the relative stability of our polarization calibrations on short time scales to ensure that the variability we observed is the result of intrinsic variations and not systematic calibration effects. Therefore, we performed our full polarimetric analysis on the phase calibrator (J2025$+$3343), grouping scans within the same bins as used for V404 Cyg when making images. 

\begin{figure*}
    \centering
    \includegraphics[width=0.80\linewidth]{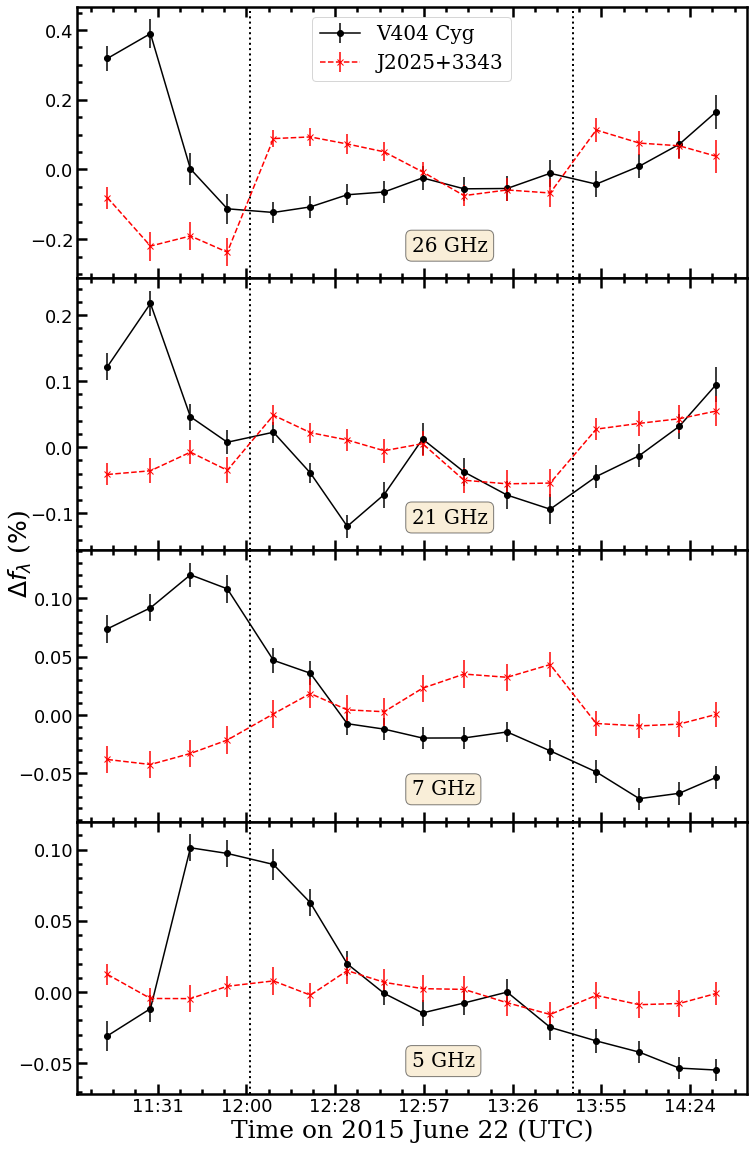}
    \caption{Temporal evolution of the linear polarization fraction residuals; 26\GHz (top), 21\GHz (2nd from the top), 7\GHz (3rd from the top), 5\GHz (bottom). The red, dashed line represents the phase calibrator J2025$+$3343, and the black, solid line, V404 Cyg. The vertical dashed lines highlight the times when the observing band switched from one sub-array to another; the jumps are the result of this transition. We can see that the linear polarization fraction evolution of J2025$+$3343 does not track V404 Cyg, and shows smaller amplitude variability in all base-bands.}
    \label{fig:phasecal_frac}
\end{figure*}

Figure \ref{fig:phasecal_frac} shows the temporal evolution of the residual rotation measure and residual observed EVPA for both V404 Cyg and J2025$+$3343. Here we define the residual as the difference between the individual time bins, and the weighted average over all time bins. Visually, J2025$+$3343 shows both a stable rotation measure ($\text{RM} \sim  -750 \radPerSqm$) and observed EVPA ($\chi_w  \sim  -35\arcdeg$). Applying the same $\chi^2$ test as discussed in Section \ref{sec:rm_pol}, neither the rotation measure ($\chi^2=27.1/12$) nor the observed EVPA ($\chi^2=2508.3/12$) is consistent with a constant value. However, the linear polarization detections of J2025$+$3343 have a much higher signal-to-noise ratio ($\text{S}/\text{N} > 200$), and, as a result, we have likely reached a systematic threshold. As such, we believe we are  underestimating the errors using an unrestricted S/N scaling (i.e., as $\text{S}/\text{N} \rightarrow \infty$, $\sigma_{\chi_w} \rightarrow 0$). The $\chi_w$ standard deviation for J2025$+$3343 is $\sim 1.4\arcdeg$, which is equal to the smallest $\chi_w$ error for V404 Cyg (${\sim} 1.4\arcdeg$) and  $\sim 1/2 $ of our median error (${\sim}  2.6\arcdeg$). Since V404 Cyg  exhibits a $\sim 30\arcdeg$ degree rotation, the systematic variability cannot be the cause of the observed evolution. 

\begin{figure*}
    \centering
    \includegraphics[width=0.95\linewidth]{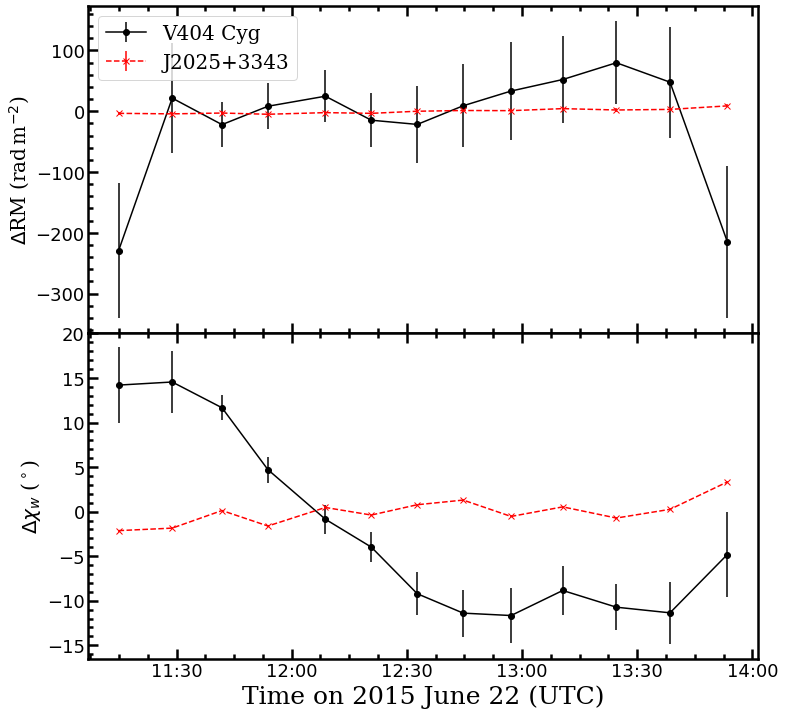}
    \caption{Temporal evolution of the residual rotation measure (top), and observed EVPA (bottom). The red, dashed line represents the phase calibrator J2025$+$3343, and the black, solid line, V404Cyg. The observed EVPA of V404 Cyg exhibits a clear evolution that is absent in J2025$+$3343.}
    \label{fig:phasecal_rm}
\end{figure*}

Figure \ref{fig:phasecal_frac} compares the temporal evolution of the linear polarization fraction, defining the ``residual'' in the same manner as in Figure \ref{fig:phasecal_rm}. The evolution of J2025$+$3343 does not track the simultaneous evolution of  V404 Cyg. The multi-band ``jumps'', at ${\sim}\,$12:00 and ${\sim}\,$13:45 UTC, correspond to a change of sub-arrays (marked by the vertical dotted lines). It is unsurprising to see some discontinuity between the two sub-arrays as each sub-array will have (slightly) different $uv$-coverage, a unique reference antenna, and, as a result, different calibration solutions. We note: (i) the bin-by-bin variability of J2025$+$3343 within a sub-array is significantly smaller than the jumps; and (ii) although obvious in J2025$+$3343, we do not observe similar jumps in our V404 Cyg data --- instead, the most significant evolution occurs absent a change of sub-array; (iii) In all four base-bands, the variability of V404 Cyg is larger than J2025$+$3343; (iv) V404 Cyg shows a common temporal evolution across base-bands, that is absent in the J2025$+$3343 data. Therefore, we are confident that the polarized detections of V404 Cyg are dominated by an intrinsic, physical evolution.

\section{Sample Images}
\begin{figure*} 
    \centering
    \includegraphics[width=0.95\linewidth]{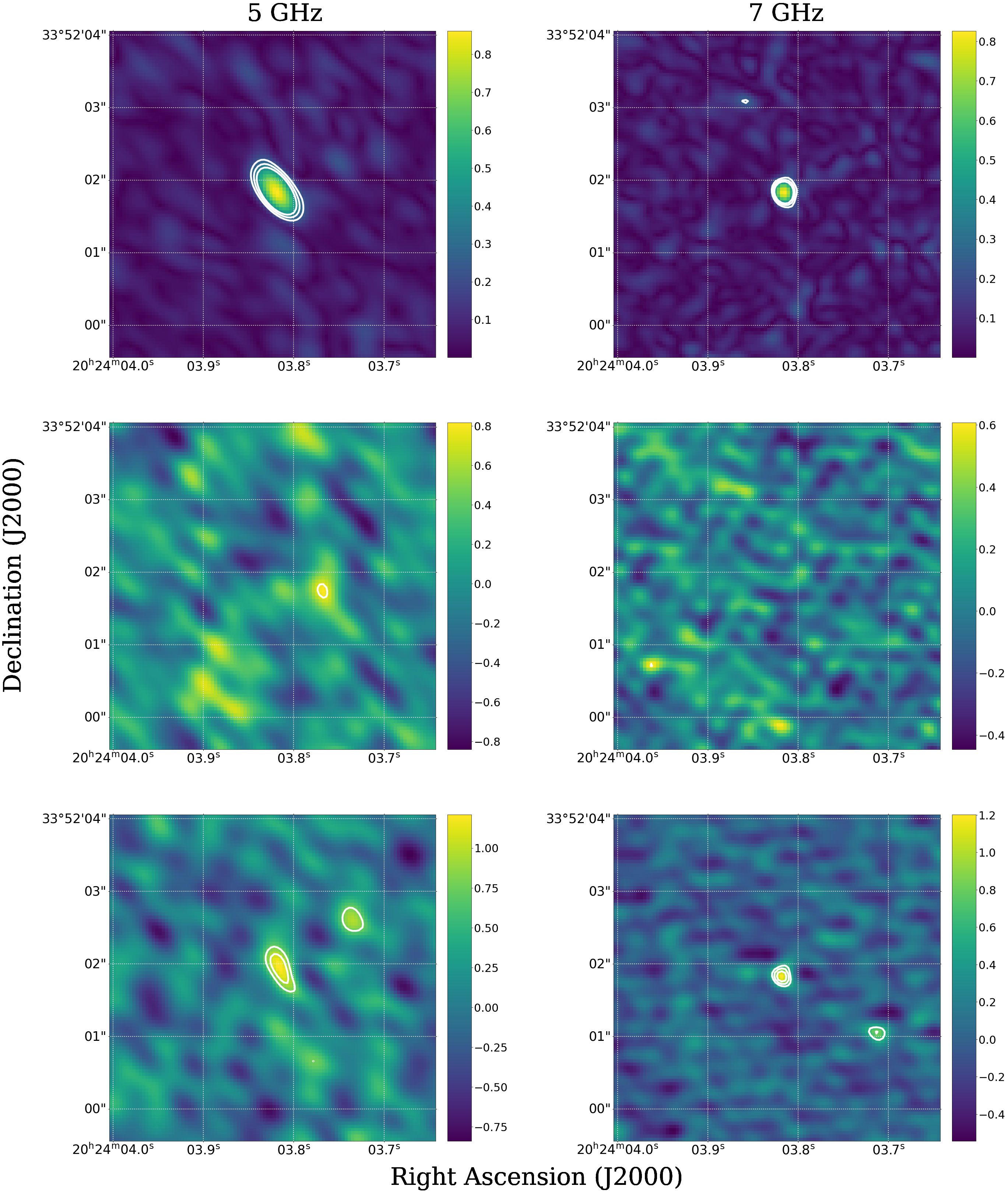}
    \caption{Sample images of the 12:09 time bin for both the 5\GHz (left) and 7\GHz (right) base-bands. (top) MFS images (i.e., $\sim\,$1\GHz bandwidths) of the linear polarization intensity ($P = \sqrt{Q^2 + U^2}$). (middle) Fine spectral resolution Stokes $Q$ images. (bottom) Fine spectral resolution Stokes $U$ images. The contours show the 3, 4, and 5$\sigma$ levels, and the color bars are in units of mJy/beam. The example images at fine spectral resolution have central frequencies of 5.209\GHz and 7.545\GHz with bandwidths of 16\MHz and 64\MHz for the 5 and 7\GHz base-bands, respectively. Stokes $Q$ and $U$ are not positive definite, and, as a result, either may appear as non-detections regardless of the strength of the detection in $P$. In this example, the source is not detected in Stokes $Q$ despite its detections in both $P$ and Stokes $U$. This behaviour motivated our use of forced aperture photometry.}
    \label{fig:example_image}
\end{figure*}

Figure \ref{fig:example_image} shows a sample set of $P$, $Q$, $U$ images at both 5 and 7\GHz. For this example, despite the clear detection in $P$ (top row), we are unable to detect the source in the Stokes $Q$ images (middle row). Due to the intrinsic variability of the source, in other time bins and frequency ranges the properties of the non-detections may change (e.g., Stokes $Q$ is detected but Stokes $U$ is not). As a result, we chose to extract the Stokes $Q$ and $U$ flux densities using forced aperture photometry.
\label{sec:appendix_image}


\section{Data Tables}
\label{sec:data_tables}

\renewcommand{\arraystretch}{1.5}

The following tables summarize the key observations: Table \ref{tab:fracpol_params} contains the polarization fraction observations; and Table \ref{tab:rm_params} contains the EVPAs and rotation measures derived from both polarimetric routines.

\begin{table*}
  \centering
  \caption{A summary of observations from the base-band integrated polarization fraction images for each base-band; t$_{ctr}$ is each time bin's central UTC time during the 22 June 2015 observations. \textit{Sig} corresponds to the probability that the detection is intrinsic to the source, and is not a calibration artifact (i.e., the significance level as defined in the text). See Section \ref{sec:obs} for definitions of the remaining parameters.}\label{tab:fracpol_params}
  \begin{tabular}{c|rccrr|rccrr}
  \Xhline{3\arrayrulewidth}
  & \multicolumn{5}{c|}{$5\,$GHz} & \multicolumn{5}{c}{$7\,$GHz} \\
  \Xhline{3\arrayrulewidth}
  $t_\text{ctr}\,$(HH:MM) & \multicolumn{1}{c}{$I_{\lambda}\,$(mJy)} & $P_{\lambda,0}\,$(mJy) & $f_\lambda\,$($\%$) & \multicolumn{1}{c}{$\frac{P_{\lambda,0}}{\sigma_{QU}}$} & \multicolumn{1}{c|}{Sig ($\%$)} & \multicolumn{1}{c}{$I_{\lambda}\,$(mJy)} & $P_{\lambda,0}\,$(mJy) & $f_\lambda\,$($\%$) & \multicolumn{1}{c}{$\frac{P_{\lambda,0}}{\sigma_{QU}}$} & \multicolumn{1}{c}{Sig ($\%$)} \\
  \Xhline{3\arrayrulewidth}

11:15 & 305.3$\,\pm\,$0.2 & 0.24$\,\pm\,$0.03 & 0.080$\,\pm\,$0.010 & 7.6 & 83 & 275.5$\,\pm\,$0.4 & 0.55$\,\pm\,$0.03 & 0.199$\,\pm\,$0.012 & 16.4 & ${>}\,99$\\
11:29 & 363.3$\,\pm\,$0.2 & 0.36$\,\pm\,$0.03 & 0.099$\,\pm\,$0.009 & 10.6 & 94 & 360.6$\,\pm\,$0.3 & 0.78$\,\pm\,$0.04 & 0.217$\,\pm\,$0.012 & 18.6 & ${>}\,99$\\
11:42 & 387.5$\,\pm\,$0.1 & 0.82$\,\pm\,$0.04 & 0.212$\,\pm\,$0.009 & 22.5 & ${>}\,99$ & 381.1$\,\pm\,$0.2 & 0.93$\,\pm\,$0.04 & 0.245$\,\pm\,$0.011 & 23.3 & ${>}\,99$\\
11:54 & 379.4$\,\pm\,$0.1 & 0.79$\,\pm\,$0.03 & 0.208$\,\pm\,$0.009 & 22.6 & ${>}\,99$ & 368.8$\,\pm\,$0.2 & 0.86$\,\pm\,$0.04 & 0.233$\,\pm\,$0.012 & 19.8 & ${>}\,99$\\
12:09 & 425.7$\,\pm\,$0.4 & 0.85$\,\pm\,$0.05 & 0.200$\,\pm\,$0.011 & 18.3 & ${>}\,99$ & 482.8$\,\pm\,$0.8 & 0.83$\,\pm\,$0.05 & 0.172$\,\pm\,$0.011 & 16.3 & ${>}\,99$\\
12:21 & 527.3$\,\pm\,$0.5 & 0.92$\,\pm\,$0.05 & 0.174$\,\pm\,$0.009 & 18.5 & ${>}\,99$ & 634.4$\,\pm\,$0.7 & 1.02$\,\pm\,$0.06 & 0.161$\,\pm\,$0.010 & 15.8 & ${>}\,99$\\
12:33 & 621.7$\,\pm\,$0.5 & 0.81$\,\pm\,$0.05 & 0.131$\,\pm\,$0.009 & 14.8 & 95 & 732.2$\,\pm\,$0.5 & 0.86$\,\pm\,$0.07 & 0.118$\,\pm\,$0.010 & 12.3 & 92\\
12:45 & 663.3$\,\pm\,$0.2 & 0.73$\,\pm\,$0.05 & 0.110$\,\pm\,$0.008 & 13.3 & 88 & 709.7$\,\pm\,$0.6 & 0.80$\,\pm\,$0.07 & 0.113$\,\pm\,$0.009 & 12.0 & 90\\
12:57 & 637.6$\,\pm\,$0.3 & 0.61$\,\pm\,$0.06 & 0.096$\,\pm\,$0.009 & 10.6 & 80 & 620.8$\,\pm\,$0.6 & 0.66$\,\pm\,$0.06 & 0.106$\,\pm\,$0.009 & 11.3 & 87\\
13:11 & 592.6$\,\pm\,$0.3 & 0.61$\,\pm\,$0.05 & 0.103$\,\pm\,$0.008 & 12.2 & 84 & 548.2$\,\pm\,$0.4 & 0.58$\,\pm\,$0.05 & 0.106$\,\pm\,$0.009 & 11.5 & 87\\
13:25 & 542.7$\,\pm\,$0.3 & 0.60$\,\pm\,$0.05 & 0.111$\,\pm\,$0.009 & 12.5 & 88 & 511.9$\,\pm\,$0.2 & 0.57$\,\pm\,$0.04 & 0.111$\,\pm\,$0.009 & 12.8 & 89\\
13:39 & 514.1$\,\pm\,$0.2 & 0.44$\,\pm\,$0.05 & 0.086$\,\pm\,$0.009 & 9.3 & 73 & 519.2$\,\pm\,$0.2 & 0.49$\,\pm\,$0.05 & 0.095$\,\pm\,$0.009 & 10.4 & 80\\
13:53 & 504.0$\,\pm\,$0.2 & 0.39$\,\pm\,$0.04 & 0.076$\,\pm\,$0.009 & 8.9 & 81 & 569.9$\,\pm\,$0.5 & 0.44$\,\pm\,$0.06 & 0.077$\,\pm\,$0.010 & 7.7 & 79\\
14:07 & 538.3$\,\pm\,$0.2 & 0.37$\,\pm\,$0.04 & 0.069$\,\pm\,$0.007 & 9.2 & 73 & 650.7$\,\pm\,$0.4 & 0.35$\,\pm\,$0.06 & 0.054$\,\pm\,$0.009 & 5.7 & 54\\
14:20 & 587.5$\,\pm\,$0.4 & 0.34$\,\pm\,$0.04 & 0.057$\,\pm\,$0.008 & 7.5 & 60 & 727.7$\,\pm\,$0.4 & 0.42$\,\pm\,$0.07 & 0.058$\,\pm\,$0.010 & 5.9 & 60\\
14:32 & 630.8$\,\pm\,$0.2 & 0.35$\,\pm\,$0.05 & 0.056$\,\pm\,$0.008 & 7.0 & 58 & 711.4$\,\pm\,$0.6 & 0.51$\,\pm\,$0.07 & 0.072$\,\pm\,$0.010 & 7.3 & 75\\
  \Xhline{3\arrayrulewidth}
  & \multicolumn{5}{c|}{$21\,$GHz} & \multicolumn{5}{c}{$26\,$GHz} \\
  \Xhline{3\arrayrulewidth}
  $t_\text{ctr}\,$(HH:MM) & \multicolumn{1}{c}{$I_{\lambda}\,$(mJy)} & $P_{\lambda,0}\,$(mJy) & $f_\lambda\,$($\%$) & \multicolumn{1}{c}{$\frac{P_{\lambda,0}}{\sigma_{QU}}$} & \multicolumn{1}{c|}{Sig ($\%$)} & \multicolumn{1}{c}{$I_{\lambda}\,$(mJy)} & $P_{\lambda,0}\,$(mJy) & $f_\lambda\,$($\%$) & \multicolumn{1}{c}{$\frac{P_{\lambda,0}}{\sigma_{QU}}$} & \multicolumn{1}{c}{Sig ($\%$)} \\
  \Xhline{3\arrayrulewidth}
11:15 & 516.8$\,\pm\,$0.9 & 2.01$\,\pm\,$0.10 & 0.388$\,\pm\,$0.020 & 19.4 & 96 & 604.3$\,\pm\,$1.0 & 4.09$\,\pm\,$0.21 & 0.677$\,\pm\,$0.035 & 19.2 & ${>}\,99$\\
11:29 & 498.1$\,\pm\,$1.0 & 2.41$\,\pm\,$0.10 & 0.484$\,\pm\,$0.019 & 25.2 & ${>}\,99$ & 482.4$\,\pm\,$1.5 & 3.61$\,\pm\,$0.20 & 0.748$\,\pm\,$0.042 & 17.8 & ${>}\,99$\\
11:42 & 457.0$\,\pm\,$0.8 & 1.43$\,\pm\,$0.09 & 0.313$\,\pm\,$0.020 & 15.8 & 88 & 510.6$\,\pm\,$1.8 & 1.84$\,\pm\,$0.23 & 0.360$\,\pm\,$0.046 & 7.8 & 80\\
11:54 & 749.3$\,\pm\,$2.0 & 2.05$\,\pm\,$0.14 & 0.274$\,\pm\,$0.018 & 15.1 & 80 & 928.3$\,\pm\,$2.5 & 2.28$\,\pm\,$0.40 & 0.246$\,\pm\,$0.043 & 5.7 & 53\\
12:09 & 1249.0$\,\pm\,$0.8 & 3.61$\,\pm\,$0.19 & 0.289$\,\pm\,$0.016 & 18.5 & 88 & 1386.3$\,\pm\,$1.0 & 3.26$\,\pm\,$0.41 & 0.235$\,\pm\,$0.029 & 8.0 & 61\\
12:21 & 1162.2$\,\pm\,$1.4 & 2.64$\,\pm\,$0.18 & 0.228$\,\pm\,$0.015 & 14.7 & 73 & 1194.7$\,\pm\,$2.1 & 3.00$\,\pm\,$0.39 & 0.251$\,\pm\,$0.032 & 7.8 & 65\\
12:33 & 859.9$\,\pm\,$1.8 & 1.26$\,\pm\,$0.15 & 0.147$\,\pm\,$0.017 & 8.5 & 42 & 781.7$\,\pm\,$2.1 & 2.23$\,\pm\,$0.24 & 0.286$\,\pm\,$0.031 & 9.4 & 75\\
12:45 & 572.5$\,\pm\,$0.8 & 1.11$\,\pm\,$0.11 & 0.195$\,\pm\,$0.020 & 9.9 & 61 & 485.1$\,\pm\,$0.7 & 1.43$\,\pm\,$0.15 & 0.294$\,\pm\,$0.032 & 9.3 & 77\\
12:57 & 468.3$\,\pm\,$0.5 & 1.31$\,\pm\,$0.12 & 0.279$\,\pm\,$0.025 & 11.4 & 86 & 409.7$\,\pm\,$0.3 & 1.37$\,\pm\,$0.14 & 0.334$\,\pm\,$0.035 & 9.5 & 85\\
13:11 & 501.1$\,\pm\,$1.0 & 1.15$\,\pm\,$0.11 & 0.229$\,\pm\,$0.021 & 10.9 & 73 & 534.4$\,\pm\,$1.9 & 1.62$\,\pm\,$0.18 & 0.303$\,\pm\,$0.034 & 9.0 & 79\\
13:25 & 851.7$\,\pm\,$2.1 & 1.65$\,\pm\,$0.18 & 0.194$\,\pm\,$0.021 & 9.2 & 61 & 1075.1$\,\pm\,$2.8 & 3.27$\,\pm\,$0.39 & 0.304$\,\pm\,$0.037 & 8.3 & 79\\
13:39 & 1223.9$\,\pm\,$1.5 & 2.11$\,\pm\,$0.28 & 0.173$\,\pm\,$0.023 & 7.7 & 53 & 1462.3$\,\pm\,$0.9 & 5.08$\,\pm\,$0.58 & 0.347$\,\pm\,$0.040 & 8.8 & 87\\
13:53 & 1231.9$\,\pm\,$0.6 & 2.74$\,\pm\,$0.22 & 0.222$\,\pm\,$0.018 & 12.3 & 66 & 1228.9$\,\pm\,$0.8 & 3.89$\,\pm\,$0.45 & 0.316$\,\pm\,$0.036 & 8.7 & 71\\
14:07 & 1259.3$\,\pm\,$0.6 & 3.20$\,\pm\,$0.22 & 0.254$\,\pm\,$0.017 & 14.5 & 75 & 1248.9$\,\pm\,$1.3 & 4.59$\,\pm\,$0.44 & 0.368$\,\pm\,$0.035 & 10.5 & 81\\
14:20 & 1021.5$\,\pm\,$2.6 & 3.04$\,\pm\,$0.20 & 0.298$\,\pm\,$0.019 & 15.5 & 86 & 913.5$\,\pm\,$3.0 & 3.93$\,\pm\,$0.36 & 0.431$\,\pm\,$0.039 & 10.9 & 90\\
14:32 & 568.6$\,\pm\,$2.0 & 2.05$\,\pm\,$0.15 & 0.361$\,\pm\,$0.027 & 13.6 & 94 & 469.2$\,\pm\,$1.9 & 2.46$\,\pm\,$0.22 & 0.524$\,\pm\,$0.048 & 11.0 & 97\\
  \Xhline{3\arrayrulewidth}
  \end{tabular}
\end{table*}

\renewcommand{\arraystretch}{1.5}

\begin{table*}
  \centering
  \caption{A summary of the RM synthesis and MCMC results, and $t_{ctr}$ adopts the same definition as used in Table \ref{tab:fracpol_params}. In this chart we've defined the S/N as the amplitude of the FDF component over the rms error across the FDF; any component with a S/N$\,{>}\,5$ was recorded, but only one time bin (at 11:29 UTC) had a secondary. The lone secondary component had a similar magnitude ($|RM|\,{\sim}\,2300{\rm\,rad\,m^{-2}}$) as the systematic errors discussed in Section~\ref{sec:rm_pol}. We do not believe this to be a real signal and have omitted this component from the table. Note that the two components with the most significant deviations from the weighted mean (${\sim}\,{-}\,100{\rm\,rad\,m^{-2}}$) are also the lowest S/N detections (S/N${\sim}\,6$).}\label{tab:rm_params}
  \begin{tabular}{c|rrrc|rccc}
  \Xhline{3\arrayrulewidth}
  & \multicolumn{4}{c|}{RM Synthesis} & \multicolumn{4}{c}{MCMC} \\
  \Xhline{3\arrayrulewidth}
  $t_\text{ctr}\,$(HH:MM) & \multicolumn{1}{c}{RM$\,$(rad$\,$m$^{-2}$)} & \multicolumn{1}{c}{$\chi_w\,$($\arcdeg$)} & \multicolumn{1}{c}{$\chi_0\,$($\arcdeg$)} & S/N & \multicolumn{1}{c}{RM$\,$(rad$\,$m$^{-2}$)} & $\chi_w\,$($\arcdeg$) & \multicolumn{2}{c}{$\chi_0\,$($\arcdeg$)}\\
  \Xhline{3\arrayrulewidth}

11:15 & $-330_{-110}^{+110}$ & $62_{-4}^{+4}$ & \phn$76_{-5}^{+5}$ & $6.8$ & $-328_{-60}^{+60}$ & $63_{-3}^{+3}$ & \multicolumn{2}{c}{$77_{-3}^{+3}$} \\
11:29 & $-79_{-90}^{+90}$ & $62_{-3}^{+3}$ & $76_{-4}^{+4}$ & $8.3$ & $-35_{-60}^{+70}$ & $64_{-3}^{+3}$ & \multicolumn{2}{c}{$78_{-3}^{+4}$} \\
11:42 & $-122_{-40}^{+40}$ & $59_{-1}^{+1}$ & $73_{-3}^{+3}$ & $20.2$ & $-127_{-30}^{+30}$ & $61_{-2}^{+1}$ & \multicolumn{2}{c}{$75_{-2}^{+2}$} \\
11:54 & $-92_{-40}^{+40}$ & $52_{-1}^{+1}$ & $66_{-3}^{+3}$ & $19.7$ & $-110_{-40}^{+40}$ & $53_{-2}^{+2}$ & \multicolumn{2}{c}{$67_{-3}^{+3}$} \\
12:09 & $-75_{-40}^{+40}$ & $46_{-2}^{+2}$ & $61_{-3}^{+3}$ & $17.5$ & $-70_{-30}^{+30}$ & $45_{-2}^{+2}$ & \multicolumn{2}{c}{$59_{-2}^{+2}$} \\
12:21 & $-115_{-50}^{+50}$ & $43_{-2}^{+2}$ & $58_{-3}^{+3}$ & $16.8$ & $-91_{-40}^{+40}$ & $41_{-2}^{+2}$ & \multicolumn{2}{c}{$56_{-3}^{+3}$} \\
12:33 & $-122_{-60}^{+60}$ & $38_{-2}^{+2}$ & $52_{-3}^{+3}$ & $11.9$ & $-100_{-50}^{+50}$ & $36_{-3}^{+2}$ & \multicolumn{2}{c}{$50_{-3}^{+3}$} \\
12:45 & $-91_{-70}^{+70}$ & $36_{-3}^{+3}$ & $50_{-3}^{+3}$ & $10.9$ & $-107_{-50}^{+50}$ & $36_{-3}^{+3}$ & \multicolumn{2}{c}{$50_{-3}^{+3}$} \\
12:57 & $-67_{-80}^{+80}$ & $36_{-3}^{+3}$ & $50_{-4}^{+4}$ & $9.2$ & $-46_{-50}^{+50}$ & $38_{-3}^{+3}$ & \multicolumn{2}{c}{$52_{-3}^{+3}$} \\
13:11 & $-48_{-70}^{+70}$ & $38_{-3}^{+3}$ & $53_{-4}^{+4}$ & $10.4$ & $-95_{-60}^{+50}$ & $40_{-3}^{+3}$ & \multicolumn{2}{c}{$54_{-4}^{+3}$} \\
13:25 & $-20_{-70}^{+70}$ & $37_{-3}^{+3}$ & $51_{-3}^{+3}$ & $11.0$ & $-76_{-50}^{+50}$ & $39_{-3}^{+3}$ & \multicolumn{2}{c}{$53_{-4}^{+4}$} \\
13:39 & $-52_{-90}^{+90}$ & $36_{-3}^{+3}$ & $50_{-4}^{+4}$ & $8.2$ & $-80_{-50}^{+50}$ & $39_{-3}^{+3}$ & \multicolumn{2}{c}{$53_{-3}^{+3}$} \\
13:53 & $-315_{-130}^{+130}$ & $43_{-5}^{+5}$ & $57_{-5}^{+5}$ & $6.0$ & $-101_{-80}^{+80}$ & $37_{-6}^{+4}$ & \multicolumn{2}{c}{$51_{-6}^{+5}$} \\
  \Xhline{3\arrayrulewidth}
  \end{tabular}
\end{table*}

\bsp	
\label{lastpage}
\end{document}